\documentclass[aps,twocolumn,letterpaper,superscriptaddress,notitlepage]{revtex4-1}
\usepackage{CJK}
\usepackage{amsmath}
\usepackage{comment}
\usepackage{eufrak}
\usepackage{natbib}
\usepackage{hyperref}
\usepackage[left=0.75in, right=0.7in, top=0.7in, bottom=1in]{geometry}
\usepackage{xcolor}
\usepackage{graphicx}
\pdfoutput=1

\begin{document}
\begin{CJK*}{GB}{}
\title{Microscopic Theory of the Magnetic Susceptibility of Insulators}
\author{Alistair H. Duff$^{1}$, Aidan Lau$^{1}$, and J.E. Sipe}
\affiliation{Department of Physics, University of Toronto, Ontario M5S 1A7, Canada}

\begin{abstract}
    We present a general theory of the magnetic susceptibility of insulators that can be extended to treat spatially varying and finite frequency fields. While there are existing results in the literature for the zero frequency response that appear to be in disagreement with each other, we show that the apparent differences between them vanish with the use of various sum rules, and that our result is in agreement with them. Although our strategy is based on the use of Wannier functions, we show that our result can be written in a ``gauge invariant" form involving Bloch functions. We can write it as the sum of terms that involve the diagonal elements of the Berry connection, and this decomposition is particularly useful in considering the limit of isolated molecules. But these contributions can be repackaged to give a form independent of those diagonal elements, which is thus generally more suitable for numerical computation. We consider an h-BN model to demonstrate the practical considerations in building a model and making calculations within this formalism.   
\end{abstract}

\maketitle
\end{CJK*}

\section{Introduction}

Understanding the magnetization of materials, and the response of the magnetization to applied fields, is a fundamental problem in condensed matter physics \cite{RothMagSus,KohnBlochMagnetic,BlountMagSus,HebbornMagSus,HebbornSondheimerEarlyDiamag}. The electronic contribution to the magnetization, which often dominates, contains both orbital contributions from the motion of the electrons 
and spin contributions from the intrinsic angular momentum of the electrons. From the ``modern theory of polarization and magnetization" comes the insight that in an extended system the orbital contribution to the magnetic moment consists not only of contributions of ``atomic-like" magnetic moments that one would expect from isolated atoms, but also of contributions of ``itinerant" magnetic moments due to the motion of electrons between sites within the solid \cite{ModernTheoryVanderbilt,Elec_Polar,Magnetization_MbI,Magnetization_PI}.

The response of the macroscopic magnetization to an applied magnetic field is characterized by the magnetic susceptibility tensor $\chi$. While it might be thought that this response would be straightforward to calculate, at least in the independent particle approximation and for applied magnetic fields that are static and uniform, identifying theoretical expressions for the magnetic susceptibility and understanding the various contributions to it has proved to be a difficult task \cite{GaoGeometricalSus,OgataMagSusI,OgataMagSusII,OgataMagSusIII,OgataMagSus2017,MatsuuraOgataPeierls}. Early attempts at understanding the magnetic susceptibility of solid state systems involved treating the conduction electrons like a free electron gas and the core electrons as effectively bound to the lattice ions \cite{PeierlsEarly,TheoryofMetals,AdamsEarlyDiamag}. The Landau-Peierls formula was discovered as a modification of the Landau diamagnetism of a free electron gas, replacing the bare electron mass with the effective mass tensor to account for the effects of the periodic potential. However, in cases where the effective mass is small and thus this diamagnetic contribution is large, interband transitions can also start to have a very large contribution, since even a weak magnetic perturbation can cause mixing of bands \cite{AdamsEarlyDiamag}.

Thus the Bloch wave functions of electrons in periodic solids are central to calculating the susceptibility; both their spectral and geometric properties play a role, and both interband and intraband matrix elements are involved.
Various Bloch function expressions for the magnetic susceptibility were obtained around the 1960s by Hebborn and Sondheimer (and Luttinger and Stiles), Kohn, Roth, and Blount \cite{HebbornMagSus,KohnBlochMagnetic,HebbornSondheimerEarlyDiamag,RothMagSus,BlountMagSus}.
From these studies emerged the general consensus that the expression should contain terms such as the Landau-Peierls susceptibility and a term that reduces to the atomic diamagnetism as the lattice spacing is taken to infinity, as well as additional corrections due to the curvature of the energy bands and interband mixing, such as the Van Vleck paramagnetism.
Attempts at comparisons between expressions derived by different workers were made at the time, but we find no definitive agreement in the literature \cite{EarlyComparisons}.

Very generally the expressions derived for the susceptibility, which involved Bloch function energies and matrix elements between Bloch functions, were considered very complicated -- to some the details were of ``repellent complexity" \cite{HebbornMagSus} -- and so attempts were made to simplify them for use in addressing real materials. A focus on the orbital motion of the electrons led to the ``Fukuyama formula" \cite{FukuyamaEarlyWork,FukuyamaBismuth}. Years later a compact expression for the susceptibility of an insulator was derived by Mauri and Louie \cite{LouieFormula}, which however involved the numerical evaluation of a double derivative of a Bloch function expression. This formula only includes the orbital contribution to the magnetic susceptibility, and also appears to be gauge-dependent in the electromagnetic sense, since the derivation depends on the choice of gauge for the vector potential. Later a compact formula for the orbital susceptibility of tight binding models was obtained \cite{TBArnaud}. To compare these formulas and test the intuitive understanding of the various contributions, a natural approach is to write all the expressions in the Bloch function basis.

Written in the Bloch basis, it was shown that the Fukuyama formula can be divided into four contributions \cite{OgataMagSusI}: the Landau-Peierls susceptibility $\chi_\text{LP}$; an interband contribution, $\chi_\text{inter}$, that reduces to the Van Vleck susceptibility in the atomic limit; a contribution from the occupied states, $\chi_\text{occ}$, that reduces  to the atomic diamagnetism from core-level electrons in the atomic limit; and a contribution from the Fermi surface, $\chi_\text{FS}$. Generalizations to include the Zeeman and spin-orbit interactions, and to allow the treatment of systems where both inversion and time-reversal symmetry are broken, lead to two new additions \cite{OgataMagSus2017}: $\chi_\text{FS-P}$, a contribution from the Fermi surface related to Pauli paramagnetism; and $\chi_\text{occ2}$, which is related to the Berry curvature of the cell periodic Bloch functions. Recently there has been particular attention paid to parts of these expressions that can be considered ``geometric," due to their connection to the Berry curvature, with interest in how quantum geometry can influence the magnetic susceptibility and be probed by it. This has been a focus of the studies by Gao et al. \cite{GaoGeometricalSus} and Piechon et al. \cite{Piechon}

Within the last decade there have been efforts to find a consensus among the different expressions for the susceptibility. Ogata and Fukuyama \cite{OgataMagSusI} showed that their work was in agreement with the earlier studies of Hebborn, Luttinger, Sondheimer and Stiles (see section II and Appendix C of Ogata and Fukuyama \cite{OgataMagSusI}), a promising connection to the early literature. Additionally, while the formula of Mauri and Louie \cite{LouieFormula} is implemented numerically, its analytic form can be identified and be seen to agree with that of Ogata \cite{OgataMagSus2017} in the insulating limit, if a symmetric gauge is chosen for the vector potential.

Yet serious disagreements persist. There are seemingly two camps that arise, and the geometric term is at the heart of the dispute. Limiting the discussion to insulators, there is a prefactor of 3/2 multiplying the geometric term in the work of Blount \cite{BlountMagSus} and Gao et al. \cite{GaoGeometricalSus} when compared to studies of Ogata \cite{OgataMagSus2017} and Piechon et al. \cite{Piechon} Suggested explanations for the discrepancy are the use of a formula derived for infinite bands in the two-band tight binding limit, and the use of a wave-packet approximation in one of the strategies \cite{Piechon,OgataMagSus2017,BlountMagSus,GaoGeometricalSus}, but ultimately the reason for the difference is unknown. As well, there are more subtle differences among some of the derivations, including the choice of the ``magnetization matrix element" and the inclusion or removal of diagonal elements of the Berry connection, and these certainly complicate comparisons. Furthermore, of all the approaches mentioned, none are positioned to generalize to a response at finite frequency. 

With this as background, in this paper: (1) we introduce a strategy for determining the magnetic susceptibility of a solid that can be generalized to finite frequency; (2) we clarify the source of the disagreements in the recent literature; and (3) we apply this formalism to a simple but realistic model to show the importance of the various contributions, in particular the geometric contribution.

Our work is based on a recent extension of a theory of the microscopic polarization and magnetization of a solid to include Zeeman and spin-orbit interactions \cite{Perry_Sipe,PerryMagneto,PerryOptical,AlistairSipe}. This approach allows for the determination of susceptibilities not from a thermodynamic framework, but instead from perturbation theory based on an electrodynamics perspective, and with generalizations to spatially and temporally varying fields possible. The method is gauge-independent in the electromagnetic sense, and provides physical insight into the different terms that contribute to susceptibilities. It employs a Wannier function basis that avoids some of the issues raised by the unbounded nature of the position operator \cite{ModernTheoryVanderbilt,SeitzSolid}.

In this paper we obtain the response of the magnetization to a uniform and constant magnetic field -- that is, we derive an expression for the magnetic susceptibility in these limits -- using this microscopic theory \cite{Perry_Sipe,AlistairSipe}. We make the independent particle approximation and consider topologically trivial insulators at zero temperature. Thus, ``free electron" terms like $\chi_\text{LP}$ will not be present, as they depend on \textbf{k} derivatives of the band filling factors. Despite our use of Wannier functions, which are not uniquely defined even if we restrict ourselves to ``maximally localized Wannier functions" \cite{MaxLocWannier,BlochBundlesMaxLoc,ExponentialLocWannier,Z2Insulator}, we show that our result can be written in terms of Bloch functions, and is independent of the ambiguity in how those functions are chosen. Thus it is ``gauge-invariant" with respect to the choice of those functions. 

We can make direct comparison to the expressions for the magnetic susceptibility already present in the literature \cite{OgataMagSusI,OgataMagSus2017}. At first glance our result looks qualitatively different than those of earlier work: there is a different definition of magnetization matrix element, and extra interband terms appear, including an additional contribution to the susceptibility that is quadratic in the Berry curvature. Yet with the use of sum rules we show that this disagreement is only apparent. Furthermore, we show how the numerical prefactor difference mentioned above is only an apparent difference when the diagonal elements of the Berry connection are treated with care. So in the course of producing a new derivation of the susceptibility, we remove the confusion in the earlier literature.  As well, we 
can write our expression in a form in which the diagonal elements of the Berry connection do not appear, and which is thus amenable to evaluation using Bloch functions found numerically.

An outline of the paper follows.

In Section \ref{MicroTheoryOutline} we outline the microscopic theory and how to apply it to the problem of finding the magnetic susceptibility. In Section \ref{SpontMag_Sect} we begin with the accepted formula for the spontaneous magnetization \cite{ModernTheoryVanderbilt}, and from it introduce an expression for the ``spontaneous magnetization matrix elements." These matrix elements are central to the magnetic susceptibility expression we obtain. 

In Section \ref{GeneralLinResponse} we show how the linear response can be partitioned into two types of contributions. We term those originating from how the magnetic field alters the populations as ``dynamical," and those originating from how the matrix elements themselves depend on the magnetic field as ``compositional." The dynamical contributions are evaluated in Section \ref{SectionDynamicalSus}, and the compositional contributions in Section \ref{SectionCompositionalSus}. 

In Section \ref{SectionTotalSus} we combine these two contributions; it is this total susceptibility that is gauge invariant. We compare our result to those of earlier studies \cite{GaoGeometricalSus,HebbornSondheimerEarlyDiamag,BlountMagSus,OgataMagSus2017}, showing that with the use of certain sum rules apparently different expressions are all equivalent. We rewrite our expression in a form involving only off-diagonal elements of the Berry connection. 

In Section \ref{SectionMolecule} we consider our total susceptibility expression in the ``molecular crystal" limit, where the solid is thought of as molecules on lattice sites but between which electrons cannot flow.  We show how the contributions simplify to the regular Van Vleck paramagnetism with the spin angular momentum and spin-orbit coupling modifications, and the Langevin atomic diamagnetism. A ``geometric" term that is present in a solid vanishes in the molecular crystal limit, as it requires matrix elements between different lattice sites.

In Section \ref{SectionModelHamiltonian} we apply this formalism to the calculation of the magnetic susceptibility of a monolayer of hexagonal boron nitride as an illustrative example. We outline two approximation schemes and investigate the effectiveness of some of the required approximations to the model.  

In Section \ref{Conclusions} we summarize our results and conclude. Some of the relevant expressions from earlier work, and details of expressions derived here, are presented in Appendices \ref{Appendix:relegatedexpressions}, \ref{Appendix:SusceptibilityCombinations} and \ref{AppendixhBN}. Details of the manipulations of some of the terms that arise are presented in an archive posting \cite{AlistairAncillaryMagSus}. 

\section{Outline of Microscopic Theory}\label{MicroTheoryOutline}

In the Heisenberg picture the dynamics of the fermionic electron field operator $\hat{\psi}(\textbf{x},t)$ is governed by 

\begin{equation}\label{EoM-Heisenberg}
    i\hbar \frac{\partial \hat{\psi}(\textbf{x},t)}{\partial t} = \mathcal{H}(\textbf{x},t)\hat{\psi}(\textbf{x},t),
\end{equation}
where $\hat{\psi}(\textbf{x},t)$ is a two-component Pauli spinor operator. We make the frozen ion and independent particle approximations, and begin by taking the electrons to be subject to a potential energy term V(\textbf{x}) that has the same periodicity as the crystal lattice, V(\textbf{x}) = V(\textbf{x}+\textbf{R}) for all Bravais lattice vectors \textbf{R}. Often the unperturbed Hamiltonian is then taken as
\begin{equation} \label{TRS}
\begin{split}
    \mathcal{H}^{0}_{TRS}(\textbf{x}) = -\frac{\hbar^2}{2m} \nabla^2 + \text{V}(\textbf{x}) + \mathcal{H}^0_{soc}(\textbf{x}),
\end{split}
\end{equation}
where spin-orbit coupling is included by using 
\begin{equation}
\label{HSOC}
    \mathcal{H}^0_{soc}(\textbf{x}) = -\frac{i\hbar^2}{4m^2c^2} \boldsymbol\sigma \cdot \nabla \text{V}(\textbf{x}) \times \nabla,
\end{equation}
where $\boldsymbol\sigma$ is the vector of Pauli matrices that act on spinor wavefunctions. We use the subscript TRS to indicate the Hamiltonian (\ref{TRS}) satisfies time-reversal symmetry. This can be broken by introducing an ``internal" static, cell-periodic vector potential $\textbf{A}_\text{static}(\textbf{x})$ \cite{Haldane_Julen,QHE_Haldane}. Its presence breaks time-reversal symmetry, but does not break the translational symmetry of the Hamiltonian, and thus Bloch's theorem can still be applied. The goal of this prescription is to allow for the possibility that the ground state eigenfunctions used in the later calculations do not satisfy time-reversal symmetry constraints, while still being of the Bloch form. This can lead to a non-zero spontaneous magnetization as well, other BZ integrals that would trivially vanish in the case of time-reversal symmetry are non-zero. In earlier studies, an effective magnetic field, nonuniform but with the periodicity of the lattice, is explicitly included for a similar purpose \cite{OgataMagSus2017}. Such a field can capture a mean-field description of exchange effects; see, e.g., Kohn and Sham \cite{KohnSham-Ex}. We do not explicitly implement that approach here, because we want to simplify the comparison of our results with earlier calculations that were done with the more basic independent particle approximation where such exchange effects are neglected.

The introduction of an applied electromagnetic field described by the scalar and vector potentials ($\phi(\textbf{x},t) , \textbf{A}(\textbf{x},t)$), together with the inclusion of the cell-periodic vector potential $\textbf{A}_\text{static}(\textbf{x})$, is done with the usual minimal coupling prescription, by taking 
\begin{equation}
    -i\hbar \nabla  \rightarrow -i\hbar\nabla-\frac{e}{c}\Big(\textbf{A}_\text{static}(\textbf{x}) + \textbf{A}(\textbf{x},t)\Big),
\end{equation}
and by including the usual scalar potential term in the Hamiltonian. We also add a Zeeman term 
\begin{equation}
    -\frac{e\hbar}{2mc} \boldsymbol\sigma \cdot \textbf{B}(\textbf{x},t),
\end{equation}
where \textbf{B}(\textbf{x},t) is the combination of both the applied and internal field. 
Setting the applied scalar and vector potentials ($\phi(\textbf{x},t) , \textbf{A}(\textbf{x},t)$) to zero one obtains the unperturbed Hamiltonian $\mathcal{H}^{(0)}(\textbf{x})$ \cite{AlistairSipe}, which extends (\ref{TRS}, \ref{HSOC}) by including $\textbf{A}_\text{static}(\textbf{x})$.

The spinor wavefunctions can be expanded in a Bloch basis
\begin{equation}
    \psi_{n\textbf{k}}(\textbf{x}) \equiv \langle \textbf{x} | \psi_{n\textbf{k}} \rangle  = \frac{1}{(2\pi)^{3/2}} e^{i\textbf{k}\cdot\textbf{x}} u_{n\textbf{k}}(\textbf{x}),
\end{equation}
where $\psi_{n\textbf{k}}(\textbf{x})$ is an eigenvector of the unperturbed Hamiltonian with energy $E_{n\textbf{k}}$, and $u_{n\textbf{k}}(\textbf{x}) \equiv \langle \textbf{x}|n\textbf{k}\rangle$ is the associated cell-periodic function. One can also define a Wannier basis as
\begin{equation}
\label{WannierFunc}
    |\alpha\textbf{R}\rangle = \sqrt{\frac{ \mathcal{V}_{uc}}{(2\pi)^3} } \int_{BZ} d\textbf{k} e^{-i\textbf{k}\cdot\textbf{R}} \sum_{n} U_{n\alpha}(\textbf{k}) |\psi_{n\textbf{k}}\rangle,
\end{equation}
where $\mathcal{V}_{uc}$ is the volume of the unit cell. Each Wannier ket and its associated Wannier function $W_{\alpha \textbf{R}}(\textbf{x)} \equiv \langle \textbf{x} | \alpha \textbf{R} \rangle$ is labeled by a type index $\alpha$ and a lattice site $\textbf{R}$ with which it is identified. 
The Bloch functions are normalized such that $\langle \psi_{n\textbf{k}}|\psi_{m\textbf{k}'} \rangle = \delta_{nm}\delta(\textbf{k}-\textbf{k}')$, and likewise $\langle \beta\textbf{R}'|\alpha\textbf{R}\rangle = \delta_{\alpha\beta} \delta_{\textbf{R}\textbf{R}'}$. For the class of insulators we consider we associate a set of Wannier functions with each isolated set of bands, such that the bands in each set may intersect among themselves but not with bands from different sets. The unitary matrix U(\textbf{k}) and the Bloch eigenvectors are chosen to be periodic over the first Brillouin zone. 

Very different Wannier functions can be constructed using the freedom in choosing the matrix U(\textbf{k}) \cite{MaxLocWannier}. Additionally, the choice of Bloch functions is not unique since one can apply an arbitrary \textbf{k}-dependent complex phase and recover equally valid eigenstates. In the following sections when we refer to quantities as being ``gauge dependent" it is in this Wannier and Bloch function sense, and when we call a quantity ``gauge invariant" it is because it does not depend upon U(\textbf{k}) or its derivatives. An example of a response tensor that does not depend on the individual phases of the Bloch functions, but is dependent on the more general unitary transformation that mixes occupied bands, is the Chern-Simons contribution to the magnetoelectric polarizability \cite{VanderbiltOMP,AlistairSipe}.

In the presence of an applied magnetic field it is useful to work with a set of adjusted Wannier functions $\bar{W}_{\alpha\textbf{R}}(\textbf{x},t)$. In the limit of a weak applied magnetic field a perturbative expansion for the adjusted Wannier functions yields
\begin{equation}
    \bar{W}_{\alpha\textbf{R}}(\textbf{x},t) = e^{i\Phi(\textbf{x},\textbf{R};t)} \chi_{\alpha\textbf{R}}(\textbf{x},t),
\end{equation}
where
\begin{equation}
\label{adjustedWannierFunctions}
\begin{split}
    &\chi_{\alpha\textbf{R}}(\textbf{x},t) = W_{\alpha\textbf{R}}(\textbf{x})
    - \frac{i}{2} \sum_{\beta\textbf{R}'} W_{\beta\textbf{R}'} 
    \\
    &\times
    \int d\textbf{z} W^*_{\beta\textbf{R}'}(\textbf{z}) \Delta(\textbf{R}',\textbf{z},\textbf{R};t) W_{\alpha\textbf{R}} (\textbf{z})+ ...
\end{split}
\end{equation}
Here
\begin{equation}
    \Phi(\textbf{x},\textbf{R};t) \equiv \frac{e}{\hbar c} \int d\textbf{w} s^i(\textbf{w};\textbf{x},\textbf{R}) A^i(\textbf{w},t) 
\end{equation}
is a generalized Peierls phase factor; the ``relator" $s^{i}(\textbf{x};\textbf{y},\textbf{R})$ is
\begin{equation}
\label{relator-s}
    s^i(\textbf{w};\textbf{x},\textbf{R}) = \int_{C(\textbf{x},\textbf{R})} dz^i \delta(\textbf{w}-\textbf{z}),
\end{equation}
where $C(\textbf{x},\textbf{R})$ specifies a path from \textbf{R} to \textbf{x}. The relators $s^{i}(\textbf{x};\textbf{y},\textbf{R})$ and $\alpha^{jk}(\textbf{x};\textbf{y},\textbf{R})$ (see (\ref{relator-alpha}) below) have been introduced and discussed previously \cite{Perry_Sipe,PerryMagneto,PerryOptical,HealyQuantum}, while the function $\Delta(\textbf{x},\textbf{y},\textbf{z};t)$ is $\Phi(\textbf{z},\textbf{x};t)+\Phi(\textbf{x},\textbf{y};t)+\Phi(\textbf{y},\textbf{z};t)$, which is simply a closed line integral of the vector potential. The rationale for 
using these adjusted Wannier functions is given in Section IIc of Duff and Sipe \cite{AlistairSipe}. 

We expand the fermionic field operator in terms of the adjusted Wannier functions and their associated fermionic creation and annihilation operators,

\begin{equation}
    \hat{\psi}(\textbf{x},t) = \sum_{\alpha,\textbf{R}} \hat{a}_{\alpha\textbf{R}}(t) \bar{W}_{\alpha\textbf{R}}(\textbf{x},t),
\end{equation}
and associated with this is a single particle density matrix (SPDM)
\begin{equation}
    \eta_{\alpha\textbf{R};\beta\textbf{R}'}(t) \equiv \langle \hat{a}^\dag_{\beta\textbf{R}'}(t) \hat{a}_{\alpha\textbf{R}}(t)\rangle e^{i\Phi(\textbf{R}',\textbf{R};t)}.
\end{equation}
As was argued earlier \cite{Perry_Sipe,AlistairSipe}, the total microscopic charge and current densities can generally be written as

\begin{equation}
\begin{split}
    &\rho(\textbf{x},t) = -\nabla\cdot\textbf{p}(\textbf{x},t) + \rho_{F}(\textbf{x},t),
    \\
    &\textbf{j}(\textbf{x},t) = \frac{\partial \textbf{p}(\textbf{x},t)}{\partial t} + c\nabla\times\textbf{m}(\textbf{x},t) + \textbf{j}_{F}(\textbf{x},t),
\end{split}
\end{equation}
where
\begin{equation}
\begin{split}
    &\rho(\textbf{x},t) \equiv \langle \hat{\rho}(\textbf{x},t)\rangle + \rho^\text{ion}(\textbf{x}),
    \\
    &\textbf{j}(\textbf{x},t) \equiv \langle \hat{\textbf{j}}(\textbf{x},t) \rangle
\end{split}
\end{equation}
with $\rho^\text{ion}(\textbf{x})$ the charge density associated with fixed ion cores, and $\langle \hat{\rho}(\textbf{x},t)\rangle$ and $\langle \hat{j}(\textbf{x},t)\rangle$ the expectation value of the charge density and current density associated with our choice of Hamiltonian. The microscopic fields $\textbf{j}_F(\textbf{x},t)$ and $\rho_F(\textbf{x},t)$ are the free current and charge densities, and $\textbf{p}(\textbf{x},t)$ and $\textbf{m}(\textbf{x},t)$ are the microscopic polarization and magnetization fields.

It is the last of these that is of interest here.  Like the other microscopic fields, the microscopic magnetization field can be decomposed into site contributions, one associated with each Bravais lattice vector \textbf{R},
\begin{equation}
    \textbf{m}(\textbf{x},t) = \sum_\textbf{R} \textbf{m}_\textbf{R}(\textbf{x},t).
\end{equation}
The site magnetization fields are further split into three contributions, 
\begin{equation}
\label{SiteMagnetizations}
    \textbf{m}_\textbf{R}(\textbf{x},t) \equiv \bar{\textbf{m}}_\textbf{R}(\textbf{x},t) + \tilde{\textbf{m}}_\textbf{R}(\textbf{x},t) + \breve{\textbf{m}}_\textbf{R}(\textbf{x},t).
\end{equation}
These are the ``atomic," ``itinerant," and ``spin" contributions respectively. The atomic magnetization is related to the site current density $\textbf{j}_\textbf{R}(\textbf{x},t)$ in the way that the magnetization of an isolated atom is related to its current density. The itinerant magnetization arises because there are corrections to this in a solid, since the sites are not isolated.
The expressions for these quantities are
\begin{equation}\label{atomicmag}
    \bar{m}^j_\textbf{R}(\textbf{x},t) \equiv \frac{1}{c} \int \alpha^{jk}(\textbf{x};\textbf{y},\textbf{R})j^{\mathfrak{p},k}_\textbf{R}(\textbf{y},t) d\textbf{y}.
\end{equation}

\begin{equation}\label{itinerantmag}
    \tilde{m}_\textbf{R}^j(\textbf{x},t) \equiv \frac{1}{c} \int \alpha^{jk}(\textbf{x};\textbf{y},\textbf{R})\tilde{j}^k_\textbf{R}(\textbf{y},t) d\textbf{y},
\end{equation}

\begin{equation}
\label{SpinMag}
\begin{split}
    \breve{\textbf{m}}_\textbf{R}(\textbf{x},t) = \frac{e\hbar}{4mc} \sum_{\alpha,\beta,\textbf{R}',\textbf{R}''} \left(\delta_{\textbf{R}\textbf{R}'} + \delta_{\textbf{R}\textbf{R}''} \right) e^{i\Delta(\textbf{R}',\textbf{x},\textbf{R}'';t)}
    \\
    \times \chi^*_{\beta\textbf{R}';\textsf{j}}(\textbf{x},t) \boldsymbol\sigma_\textsf{ji} \chi_{\alpha\textbf{R}'';\textsf{i}}(\textbf{x},t) \eta_{\alpha\textbf{R}'';\beta\textbf{R}'}(t).
\end{split}
\end{equation}
where the definitions of $\textbf{j}^{\mathfrak{p}}_\textbf{R}(\textbf{x},t)$ and $\tilde{\textbf{j}}_\textbf{R}(\textbf{x},t)$ can be found in Appendix \ref{Appendix:relegatedexpressions}. The subscript $\textsf{i},\textsf{j}$ indices denote spinor components, and as usual repeated indices are summed over. The relator $\alpha^{jk}(\textbf{w};\textbf{x},\textbf{y})$ is defined as \cite{Perry_Sipe,PerryMagneto,PerryOptical,HealyQuantum}
\begin{equation}
\label{relator-alpha}
    \alpha^{jk}(\textbf{w};\textbf{x},\textbf{y}) = \epsilon^{jmn} \int_{C(\textbf{x},\textbf{y})} dz^m \frac{\partial z^n}{\partial x^k} \delta(\textbf{w}-\textbf{z}).
\end{equation}

Turning now to the problem at hand, assuming
uniform and constant fields the macroscopic magnetization is obtained by spatially averaging the microscopic magnetization of a site,
(\ref{atomicmag}, \ref{itinerantmag}, \ref{SpinMag})
\begin{equation}
\label{MacroM}
    \textbf{M} = \frac{1}{\mathcal{V}_{uc}} \int d\textbf{x} \textbf{m}_\textbf{R}(\textbf{x}).
\end{equation}

The SPDM is a central object in calculating this quantity. For a topologically trivial insulator, before any fields are applied the SPDM is given by
\begin{equation}
    \label{eta_nought}
    \eta^{(0)}_{\alpha\textbf{R}'';\beta\textbf{R}'} = f_{\alpha} \delta_{\alpha\beta} \delta_{\textbf{R}'\textbf{R}''},
\end{equation}
where $f_{\alpha}=1$ if the Wannier function of type $\alpha$ is associated with the filled bands, and $f_{\alpha}=0$ if it is associated with the empty bands. The first order response of the SPDM to a uniform, static magnetic field is given by \cite{PerryMagneto}

\begin{widetext}

\begin{equation}
\label{etaB}
\begin{split}
    \eta^{(B)}_{\alpha\textbf{R};\beta\textbf{R}'} = \frac{e}{4\hbar c} \epsilon^{lab} B^l \frac{\mathcal{V}_{uc}}{(2\pi)^3} \sum_{mn} f_{nm} \int d\textbf{k} e^{i\textbf{k}\cdot(\textbf{R}-\textbf{R}')} U^\dag_{\alpha m}(\textbf{k}) \mathcal{B}_{mn}^{ab}(\textbf{k}) U_{n\beta}(\textbf{k})
    \\
    + \frac{e}{4\hbar c} \epsilon^{lab} \frac{\mathcal{V}_{uc}}{(2\pi)^3} B^l \sum_{mn} f_{nm} \int d\textbf{k} e^{i\textbf{k}\cdot(\textbf{R}-\textbf{R}')} \left( \partial_a U^\dag_{\alpha m}(\textbf{k}) U_{n\beta}(\textbf{k}) -U^\dag_{\alpha m}(\textbf{k}) \partial_a U_{n\beta} \right) \xi^b_{mn}(\textbf{k})
    \\
    +\frac{\mathcal{V}_{uc}}{(2\pi)^3} \frac{e}{mc} B^l \sum_{mn} f_{nm} \int_{BZ} d\textbf{k} e^{i\textbf{k}\cdot(\textbf{R}-\textbf{R}')}
    \frac{ U^\dag_{\alpha m} S^l_{mn}  U_{n\beta}}{E_{m\textbf{k}}-E_{n\textbf{k}}}
\end{split}
\end{equation}

where 

\begin{equation}
    \mathcal{B}^{ab}_{mn}(\textbf{k}) = i\sum_s \left(\frac{E_{s\textbf{k}} - E_{n\textbf{k}}}{E_{m\textbf{k}}-E_{n\textbf{k}}} + \frac{E_{s\textbf{k}} - E_{m\textbf{k}}}{E_{m\textbf{k}} - E_{n\textbf{k}}}\right) \xi^a_{ms}(\textbf{k}) \xi^b_{sn}(\textbf{k}) - 2 \frac{ \partial_a E_{m\textbf{k}} + \partial_a E_{n\textbf{k}}}{E_{m\textbf{k}} - E_{n\textbf{k}}} \xi^b_{mn}(\textbf{k}).
\end{equation}
\end{widetext}
We set $f_{m}=0,1$ if band $m$ is empty or filled respectively, and $f_{nm} \equiv f_{n}-f_{m}$. Here
\begin{equation}
\label{Berry connection}
    \xi^a_{mn} \equiv i(m\textbf{k}| \partial_a n\textbf{k})
\end{equation}
is the non-Abelian Berry connection, and
\begin{equation}
\label{SpinConnection}
    S^a_{mn} \equiv (m\textbf{k}|\frac{\hbar}{2}\sigma^a n\textbf{k}),
\end{equation}
are the spin matrix elements, where
\begin{equation}
\label{unitcellNormalization}
    (g|h) \equiv \frac{1}{\mathcal{V}_{uc}} \int_{\mathcal{V}_{uc}} d\textbf{x} g^\dag(\textbf{x}) h(\textbf{x}),
\end{equation}
and $\partial_a$ is shorthand for ${\partial}/{\partial k^a}$. The cell periodic Bloch functions obey the orthogonality condition $(m\textbf{k}|n\textbf{k}) = \delta_{mn}$. We use the notation of equation (\ref{unitcellNormalization}) since the normalization condition we have chosen for the Bloch functions (below equation (\ref{WannierFunc})) then follows from the choice of normalization for the cell-periodic part.  

\section{Spontaneous Magnetization Matrix Elements}\label{SpontMag_Sect}

The ``modern theory of polarization and magnetization" \cite{ModernTheoryVanderbilt} provides an expression for the spontaneous orbital magnetization; the result of our own microscopic theory \cite{Perry_Sipe} is in agreement with that expression.
In our approach it follows from evaluating the microscopic magnetization contributions (\ref{atomicmag}, \ref{itinerantmag}, \ref{SpinMag}) using the unperturbed SPDM (\ref{eta_nought}), and putting them in the expression (\ref{MacroM}) for the macroscopic magnetization. The atomic-like and itinerant magnetizations are individually gauge-dependent but their sum, the orbital magnetization, is gauge-invariant 
and can be written as

\begin{equation}
\label{groundstateMorb}
\begin{split}
    M^{l,(0)}_{\text{Orb}} = \frac{ie}{2\hbar c} \epsilon^{lab} \sum_{ms} f_{m} \int \frac{d\textbf{k}}{(2\pi)^3}  (E_{m\textbf{k}}+E_{s\textbf{k}})\xi^a_{ms}\xi^b_{sm}.
\end{split}
\end{equation}
While in earlier work \cite{GaoGeometricalSus,RothMagSus} the bands were assumed to be nondegenerate, here we allow for degeneracies among the filled valence bands and among the empty conduction bands. It is then worth noting that the $s=m$ terms in the above sum make no contribution, so there are no diagonal elements of the Berry connection in the theoretical expression for the ground state orbital magnetization. Indeed, there are only contributions to the magnetization from the bands $s$ that are not among the filled valence bands, for if $m_1$ and $m_2$ are two of the filled valence bands, then the contribution to (\ref{groundstateMorb}) from $m=m_1$ and $s=m_2$ will cancel that from $m=m_2$ and $s=m_1$.

The spin contribution to the spontaneous magnetization can also be written as a single integral over the Brillouin zone as

\begin{equation}
    M^{l,(0)}_\text{Spin} = \frac{e}{ mc} \sum_n f_n \int \frac{d\textbf{k}}{(2\pi)^3} S_{nn}.
\end{equation}

We can go further and introduce spontaneous magnetization matrix elements. Written such that the matrix is explicitly Hermitian, they are 

\begin{widetext}
\begin{equation}
\label{SpontMagEq}
    M^l_{mn} = \epsilon^{lab} \frac{e}{4 c}\Bigg(\sum_{s} \Big(v^b_{ms}\xi^a_{sn} +\xi^a_{ms}v^b_{sn}\Big) + \frac{1}{\hbar} \partial_b(E_{m\textbf{k}} + E_{n\textbf{k}}) \xi^a_{mn}\Bigg) + \frac{e}{mc} S^l_{mn},
\end{equation}
\end{widetext}
where the velocity matrix elements are defined as
\begin{equation}
\label{velocity}
\begin{split}
    v^i_{mn} \equiv& \int d\textbf{x} \psi^\dag_{m\textbf{k}}(\textbf{x})\Big( \frac{\mathfrak{p}^i(\textbf{x})}{m} + \epsilon^{ijk}\frac{\hbar}{4m^2c^2} \sigma^l \frac{\partial \text{V}(\textbf{x})}{\partial x^m} \Big) \psi_{n\textbf{k}}(\textbf{x})
    \\
    =  &  \delta_{mn} \frac{1}{\hbar} \partial_i E_{n\textbf{k}} + \frac{i}{\hbar} (E_{m\textbf{k}}-E_{n\textbf{k}})\xi^i_{mn},
\end{split}
\end{equation}
where here
\begin{equation}
\label{pscriptstatic}
 \mathfrak{p}(\textbf{x})=-i\hbar\nabla-\frac{e}{c}\textbf{A}_\text{static}(\textbf{x}).
\end{equation}
Note that in general the equal energy elements of the Berry connection will make a contribution to equation (\ref{SpontMagEq}), where by the ``equal energy elements" of the Berry connection  $\xi^i_{mn}(\textbf{k})$ we mean those for which $E_{m\textbf{k}}=E_{n\textbf{k}}$. For a general matrix element at a given $\textbf{k}$  (like $M^l_{mn}(\textbf{k})$) we also make the distinction between the case in which band indices $m$ and $n$ label bands that satisfy $E_{m\textbf{k}}=E_{n\textbf{k}}$, -- in the non-degenerate case these would be simply the diagonal matrix elements -- and the case in which they do not. Thus we will refer to matrix elements as being either equal energy elements in this sense, or as being ``distinct energy elements".

The unperturbed macroscopic magnetization $M^{l,(0)}_\text{Orb}+M^{l,(0)}_\text{Spin}$ can be written as  
the trace of the matrix defined by $M^{l}_{mn}$ over the filled states, and integrated over the Brillouin zone, \begin{equation}
\label{trace_formula}
M^{l,(0)}_\text{Orb}+M^{l,(0)}_\text{Spin} = \sum_n f_n \int \frac{d\textbf{k}}{(2\pi)^3} M^{l}_{nn}(\textbf{k}).
\end{equation}
To get the orbital contribution into the form shown in (\ref{groundstateMorb}) requires an additional integration by parts, and only then can it be seen that there is no dependence on the equal energy elements of the Berry connection. Nonetheless, (\ref{SpontMagEq}) is a natural choice for the spontaneous magnetization matrix elements for two reasons: First, the orbital part is decomposed into an ``atomic-like" contribution that is effectively a ``position cross velocity" matrix element (angular momentum), and an ``itinerant" contribution that arises due to a deviation from flat bands. Second, this magnetization matrix element appears in the expression for the magnetic susceptibility. In fact, the SPDM response to a magnetic field can be written involving the matrix element $M^{l}_{mn}$ divided by the energy difference of the bands, 
\begin{equation}
\begin{split}
    &\eta^{(B)}_{\alpha\textbf{R};\beta\textbf{R}'} =
    \\
    &B^l \mathcal{V}_{uc} \sum_{mn} f_{nm} \int_{BZ} \frac{d\textbf{k}}{(2\pi)^3} e^{i\textbf{k}\cdot(\textbf{R}-\textbf{R}')}
    \times\Bigg[ \frac{ U^\dag_{\alpha m} M^l_{mn} U_{n\beta}}{E_{m \textbf{k}} -E_{n \textbf{k}} 
    }
    \\
    &+ \frac{e}{4\hbar c} \epsilon^{lab} \xi^b_{mn} \Big( \partial_a U^\dag_{\alpha m} U_{n\beta} - U^\dag_{\alpha m} \partial_a U_{n\beta} \Big) \Bigg]. 
\end{split}
\end{equation}
To avoid any possible confusion, we note that in other studies \cite{OgataMagSus2017,OgataMagSusI,OgataMagSusII,OgataMagSusIII} a \textit{different} spontaneous magnetization matrix appears in expressions for the magnetic susceptibility. The trace of that different spontaneous magnetization matrix over filled states, and integrated over the Brillouin zone, leads to 
the same result for $M^{l,(0)}_\text{Orb}+M^{l,(0)}_\text{Spin}$, but the off-diagonal elements of that different matrix are such that it is 
not Hermitian. Written in our notation the alternative matrix element is (using a stylized font to distinguish it from ours), 
\begin{equation}
\label{MagOgata}
\begin{split}
    \mathcal{M}^{l}_{nm} = \frac{e}{2c} \epsilon^{lab} \Big( \sum_{l}  \xi^a_{nl} v^b_{lm} + \frac{1}{\hbar} \partial_b E_{n\textbf{k}} \xi^a_{nm} \Big) + \frac{e}{mc} S^l_{nm} .
\end{split}
\end{equation}
Using the ``general effective mass tensor sum rule"
\begin{equation}
\label{effectivemasstensor}
\begin{split}
    \partial_j v^i_{nm} = \frac{\hbar}{m} \delta_{ij} \delta_{nm} + i\sum_{l} \Big( \xi^j_{nl}v^i_{lm} - i v^i_{nl}\xi^j_{lm} \Big),
\end{split}
\end{equation}
a derivation of which can be found in section VI of supplemental material \cite{AlistairAncillaryMagSus}, we can write our spontaneous magnetization matrix element in terms of (\ref{MagOgata}), 
\begin{equation}
\label{SpontMagRelation}
    M^l_{nm} = \mathcal{M}^l_{nm} - \frac{e}{4\hbar c} (E_{n\textbf{k}}-E_{m\textbf{k}}) \Omega^l_{nm}, 
\end{equation}
where ${\bf{\Omega}}_{nm}$ is the curl of the Berry connection,
\begin{equation}
\label{nonAbelian}
\Omega^l_{nm}=\epsilon^{lab} \partial_a \xi^b_{nm}. 
\end{equation}
The definition of the spontaneous magnetization matrix, equation (\ref{MagOgata}), 
disagrees with that used in yet other earlier papers \cite{GaoGeometricalSus,BlountMagSus}, where the terms that
contain the equal energy matrix elements $\xi^a_{nn'}$ are explicitly removed in both the equal and distinct energy matrix elements of expression (\ref{MagOgata}) leading to a ``purified" matrix denoted by an overset ring, with elements

\begin{equation}
\label{PureMnm}
\begin{split}
    \mathring{\mathcal{M}}^l_{nm} = \frac{e}{2c} \epsilon^{lab} \Big(\sum_{l\neq n} \xi^a_{nl} v^b_{lm} + \frac{1}{\hbar} \partial_b E_{n\textbf{k}} \xi^a_{nm} \Big) + \frac{e}{mc} S^l_{nm}
\end{split}
\end{equation}
for $n$ and $m$ labelling distinct energy bands, and
\begin{equation}\label{PureMnn}
\begin{split}
    \mathring{\mathcal{M}}^l_{nn'} = \frac{e}{2c} \epsilon^{lab} \sum_{l\neq n} \xi^a_{nl}v^b_{ln'} + \frac{e}{mc} S^l_{nn'}, 
\end{split}
\end{equation}
when $n$ and $n'$ are equal energy bands. 
When we use the notation $l\neq n$ we mean, at a given 
\textbf{k}, the sum $l$ is only over distinct energy bands. This is an extension of earlier work that only considered non-degenerate bands \cite{RothMagSus,GaoGeometricalSus}.

When we later compare our results with those of others we will see that the choice of spontaneous magnetization matrix,
along with other differences that arise, will
cancel in the final expression for the susceptibility.

\section{General form of the linear response} 
\label{GeneralLinResponse}

We now turn to the calculation of the response to a static and uniform magnetic field. In our formalism there are two distinct types
of contributions to response tensors that arise, \textit{dynamical} and \textit{compositional}. To illustrate this, consider 
the response of the site spin magnetic dipole moments 
to an applied magnetic field. We begin by writing the expression (\ref{SpinMag}) for the spin contribution to the site magnetization as the trace over the matrix product of the site spin matrix and the single particle density matrix in the adjusted Wannier function basis,
\begin{equation}
\label{SpinMag-MatrixProduct}
\begin{split}
    \breve{m}_\textbf{R}(\textbf{x}) = \sum_{\alpha\beta\textbf{R}'\textbf{R}''} \breve{m}_{\beta\textbf{R}';\alpha\textbf{R}''}(\textbf{x},\textbf{R}) \eta_{\alpha\textbf{R}'';\beta\textbf{R}'},
\end{split}
\end{equation}
where we have defined
\begin{equation}
\label{spinsitematrixele}
\begin{split}
    \breve{\textbf{m}}_{\beta\textbf{R}';\alpha\textbf{R}''}(\textbf{x},\textbf{R}) = \frac{e\hbar}{4mc} \Big( \delta_{\textbf{R}\textbf{R}'} + \delta_{\textbf{R}\textbf{R}''}\Big) e^{i\Delta(\textbf{R}',\textbf{x},\textbf{R}'')}
    \\
    \times \chi^\dag_{\beta\textbf{R}'}(\textbf{x}) \boldsymbol\sigma \chi_{\alpha\textbf{R}''} (\textbf{x}). 
\end{split}
\end{equation}
Both these site spin matrix elements, and the components of the SPDM, depend on the magnetic field. Thus one contribution to the linear response is obtained by taking the SPDM elements to first order and the site matrix elements to zeroth order; it is dubbed the \textit{dynamical} contribution. The second contribution is obtained by taking the SPDM elements to zeroth order and the site matrix elements to first order; it is dubbed the \textit{compositional} contribution. 

The dynamical contribution is 
\begin{equation}
\begin{split}
    \breve{\textbf{m}}_{\textbf{R}\text{dyn}}(\textbf{x}) = \sum_{\alpha\beta\textbf{R}'\textbf{R}''} \breve{\textbf{m}}^{(0)}_{\beta\textbf{R}';\alpha\textbf{R}''}(\textbf{x},\textbf{R}) \eta^{(B)}_{\alpha\textbf{R}'';\beta\textbf{R}'},
\end{split}
\end{equation}
where the spin matrix elements $\breve{\textbf{m}}^{(0)}_{\beta\textbf{R}';\alpha\textbf{R}''}(\textbf{x},\textbf{R})$ involve the unperturbed Wannier functions (\ref{WannierFunc}).
The compositional contribution is 
\begin{equation}
    \breve{\textbf{m}}_{\textbf{R}\text{comp}}(\textbf{x}) = \sum_{\alpha\textbf{R}'} f_{\alpha} \breve{\textbf{m}}^{(B)}_{\alpha\textbf{R}';\alpha\textbf{R}'}(\textbf{x},\textbf{R})
\end{equation}
where we have used the ground state SPDM (\ref{eta_nought}), and $\breve{\textbf{m}}^{(B)}_{\alpha\textbf{R}';\alpha\textbf{R}'}(\textbf{x},\textbf{R})$ will depend on the magnetic field through the adjusted Wannier functions, equation (\ref{adjustedWannierFunctions}), or expanding the exponential in equation (\ref{spinsitematrixele}) to first order in $\Delta(\textbf{R}',\textbf{x},\textbf{R}'')$. There can be an additional compositional contribution if
an operator depends on the magnetic field, as does the velocity operator. In particular, in calculating the compositional contribution the expression (\ref{pscriptstatic}) for $\mathfrak{p}^i(\textbf{x})$, used in the first line of (\ref{velocity}), is replaced by
\begin{equation}
\label{pscriptdynamic}
 \mathfrak{p}^i(\textbf{x})\rightarrow-i\hbar\nabla-\frac{e}{c}\Big(\textbf{A}_\text{static}(\textbf{x}) + \textbf{A}(\textbf{x},t)\Big),
\end{equation}
which results in an additional term appearing in the second line of (\ref{velocity}).

The other two constituents of the site magnetization (see Eq. (\ref{SiteMagnetizations})) also admit of a decomposition into dynamical and compositional contributions. So then does the linear response of the macroscopic magnetization (\ref{MacroM}), and the resulting susceptibility, 
\begin{equation}
\label{chidef}
    \chi^{il} = \frac{\partial M^i}{\partial B^l}\Big|_{\textbf{B} \rightarrow 0}
\end{equation}
as well, 
\begin{equation}
\label{dypluscomp}
\chi^{il}=\chi^{il}_{\text{dyn}}+\chi^{il}_{\text{comp}}.
\end{equation}
We identify those contributions in the sections below. 
The details of deriving the atomic, itinerant, and spin contributions - both dynamical and compositional - are shown in section III of supplemental material \cite{AlistairAncillaryMagSus}.

\section{Dynamical contributions to the susceptibility
}\label{SectionDynamicalSus}

The contributions to $\chi^{il}_{\text{dyn}}$ 
arise from implementing the expression
(\ref{etaB}) for the change in the 
single particle density matrix that is 
first order in the magnetic field in the 
equations (\ref{atomicmag}, 
\ref{itinerantmag}, \ref{SpinMag}) for 
the three contributions to the site 
magnetization (\ref{SiteMagnetizations}), taking the site quantity 
matrix elements to zeroth order in the 
magnetic field, and then using the expression 
(\ref{MacroM}) for the macroscopic magnetization to identify the contributions to it that 
result. In Appendix 
\ref{Appendix:SusceptibilityCombinations}
we construct 
$\chi^{il}_{\text{dyn}}$ from those three contributions (\ref{atomicmag}, 
\ref{itinerantmag}, \ref{SpinMag})  
to the site magnetization, labeling 
them $\bar{\chi}^{il}_\text{dyn}$, 
$\tilde{\chi}^{il}_\text{dyn}$, and 
$\breve{\chi}^{il}_\text{dyn}$ 
respectively. We find that their sum can be written as 

\begin{equation}
\label{dynamical}
\begin{split}
    \chi^{il}_\text{dyn}=
    \chi^{il}_\text{VV}
    +X^{il}_\text{dyn},
\end{split}    
\end{equation}
where
\begin{equation}
\label{ChiVanVleck}
\begin{split}
    \chi^{il}_\text{VV} = \sum_{m\neq n} f_{nm} \int_{BZ} \frac{d\textbf{k}}{(2\pi)^3} \frac{ M^i_{nm} M^l_{mn} }{E_{m\textbf{k}}-E_{n\textbf{k}}}  
\end{split}
\end{equation}
is what we call the ``Van Vleck contribution" as it is a generalization of the atomic Van Vleck paramagnetism to insulators that includes spin effects; we examine this further in Section \ref{SectionMolecule}. It is very similar to the ``interband" contribution $\chi^{il}_\text{inter}$ identified by Ogata 
\cite{OgataMagSus2017}, but with our definition of the magnetization matrix element. Contributions from all the terms $\bar{\chi}^{il}_\text{dyn}$, 
$\tilde{\chi}^{il}_\text{dyn}$, 
$\breve{\chi}^{il}_\text{dyn}$ are assembled into the magnetization matrix elements that appear in (\ref{dynamical}). The first part of $M^{il}_{nm}$ (\ref{SpontMagEq}), which can be thought of as a naive cross product of position with velocity, comes from the atomic contribution.  The additional partial on energy terms that appear in equation (\ref{SpontMagEq}) arise from the itinerant magnetic dipole contribution, as would be expected since they vanish in the flat band limit in which the sites are isolated. The third term is the spin contribution to the spontaneous magnetization. Note that the Van Vleck term (\ref{ChiVanVleck}) is what one would naively expect for the linear response of the magnetization, as characterized by a matrix $M^{il}_{nm}$ at each $\textbf{k}$, under the action of a Hamiltonian involving the dot product of the magnetization at each $\textbf{k}$ and the magnetic field. 

Unlike $\chi^{il}_\text{VV}$, which only involves terms that follow from the Bloch functions and the band energies, the term $X^{il}_\text{dyn}$ also depends on the unitary matrices that link the Bloch functions to the ELWFs and their derivatives. The functional form is shown in Appendix \ref{Appendix:SusceptibilityCombinations}. We return to it below. An example of its form when a specific gauge-transformation is chosen is shown in section V of the supplemental material \cite{AlistairAncillaryMagSus}.

\section{Compositional contributions to 
the susceptibility} \label{SectionCompositionalSus}

The compositional contributions to $\chi^{il}$ (\ref{chidef}) arise from using the unperturbed single particle density matrix (\ref{eta_nought}) in the expressions (\ref{atomicmag}, 
\ref{itinerantmag}, \ref{SpinMag}) for 
the atomic, itinerant, and spin compositional contributions to the site magnetization, keeping the parts of the resulting expressions linear in the magnetic field, and using the expression (\ref{MacroM}) for the macroscopic magnetization to identify the compositional susceptibility $\chi^{il}_\text{comp}$. Details are shown in section III of the supplemental material \cite{AlistairAncillaryMagSus}. We can organize some of the terms that result into two sets, in such a way as to make them comparable to theoretical expressions in the literature. 

The first set involves contributions from the atomic and itinerant site magnetizations (\ref{atomicmag}, 
\ref{itinerantmag}), and is given by

\begin{equation}
\label{ChiOcc}
\begin{split}
    \chi^{il}_\text{occ} = \frac{e^2}{4mc^2} \epsilon^{iab}\epsilon^{lcd} \sum_{nm} f_{n} \int_{BZ} \frac{d\textbf{k}}{(2\pi)^3} \text{Re}\Bigg[
    \delta_{bc} \xi^a_{nm} \xi^d_{mn}
    \\
    - \frac{m}{\hbar^2} \xi^a_{nm}\xi^d_{mn} \partial_b\partial_c E_{n\textbf{k}}
    \Bigg].
\end{split}
\end{equation}
The name ``occupied" here at first glance does not appear appropriate, however, using the completeness relation of the Bloch functions it can be shown that
\begin{equation}
    \sum_{m} \xi^a_{nm}\xi^d_{mn} = (\partial_a u_{n\textbf{k}}| \partial_d u_{n\textbf{k}}),
\end{equation}
and thus (\ref{ChiOcc}) can be written so as not to require a sum over all states. 

The second term on the right hand side of equation (\ref{ChiOcc}) is an itinerant contribution. Involving a second derivative of the band energies, it vanishes in the limit of flat-bands and is associated with the motion of electrons between different sites. 

The first term on the right side of equation (\ref{ChiOcc}) has a correspondence to the atomic diamagnetism, and as would be expected arises from the atomic magnetization. In a free energy approach, starting from the minimal coupling Hamiltonian with a magnetic field in the $z$ direction, and choosing a symmetric gauge for the vector potential, one gets a term in the Hamiltonian proportional to $(x^2+y^2)B_z^2$. The first order correction to the energy is already second order in the magnetic field, so can be considered a contribution to the total susceptibility. This term also has an appealing semi-classical interpretation, where if a magnetic field is applied perpendicular to the plane of the orbit of an electron, the electron will speed up or slow down so that its magnetic moment opposes the direction of the applied magnetic field \cite{Griffiths}. If instead the magnetic field is at an angle to the the orbit, Larmor precession results, and the precession can be interpreted as an effective magnetic moment. In either case the change in the magnetic moment due to the magnetic field is 

\begin{equation}
    \delta\textbf{m} = -\frac{e^2 R^2}{4m} \textbf{B},
\end{equation}
where $R$ is the orbital radius. Moving to a more quantum mechanical picture the radius squared is replaced by the second moment of the charge density. And still going further, extending to the necessary quantum mechanical expressions for a band insulator, one obtains the first term on the right hand side of equation (\ref{ChiOcc}). 

Note that the bare electron mass term $1/m$ that appears in equation (\ref{ChiOcc}) arises from the use of the Schr{\"o}dinger Hamiltonian as the starting point of the derivation; the velocity operator associated with this choice of Hamiltonian is altered by a magnetic field. However, in an effective tight binding Hamiltonian calculation, or even in a band structure calculation, the number of bands must be truncated, and therefore  the effective mass sum rule, equation (\ref{effectivemasstensor}), will be violated to some degree. To circumvent this, the first term on the right side of equation (\ref{effectivemasstensor}) can be replaced by the Hessian matrix \cite{GaoFOP}. In some of the manipulations we make to prove gauge-invariance and show agreement between the different ways of writing the magnetic susceptibility, this sum rule must be used. Thus numerical agreement may not be assured. To obtain a consistent theory when applied to an effective lattice Hamiltonian, in both Eq. (\ref{effectivemasstensor}) and (\ref{ChiOcc}) the Hessian matrix substitution should be employed in place of the bare electron mass. 

A second set of terms in the compositional contributions to the susceptibility lead to a contribution that depends on the curl of the non-Abelian Berry connection. We call this a ``geometric" contribution,
\begin{widetext}
\begin{equation}
\begin{split}\label{ChiGeometric}
    \chi^{il}_\text{geo} = 
    -\frac{e}{2\hbar c} \sum_{nm} f_{n} \int_{BZ} \frac{d\textbf{k}}{(2\pi)^3}
    \text{Re}\Bigg[ 
    \Omega^i_{nm} \Big( M^l_{mn} + \frac{e}{8\hbar c} (E_{n\textbf{k}}-E_{m\textbf{k}})\Omega^l_{mn}) 
    \\
    + \Big(M^i_{nm} + \frac{e}{8\hbar c}(E_{n\textbf{k}}-E_{m\textbf{k}})\Omega^i_{nm} \Big) \Omega^l_{mn} 
    \Bigg].
\end{split}
\end{equation}
\end{widetext}
Note that this is different from the ``geometric" contribution found from other methods \cite{OgataMagSus2017,GaoGeometricalSus,BlountMagSus} as it contains interband matrix elements and 
a contribution that is quadratic in $\boldsymbol{\Omega}$ (recall Eq. (\ref{nonAbelian})). Arising purely from compositional contributions, it is of the form of a matrix multiplication of $\boldsymbol{\Omega}$ and a modified spontaneous magnetization matrix element.  

Similar to the nature of the terms constituting the dynamical contribution $\chi^{il}_{\text{dyn}}$ (\ref{dynamical}), there is an additional contribution $X^{il}_{\text{comp}}$ to the compositional contribution that depends not just on the Bloch functions and band energies, but also on the choice of Wannier functions. This term is discussed in Appendix \ref{Appendix:SusceptibilityCombinations}.  In all, then, the compositional contribution to the susceptibility can be written as
\begin{equation}
    \chi^{il}_\text{comp} = \chi^{il}_\text{occ} + \chi^{il}_\text{geo} + X^{il}_\text{comp}. 
\end{equation}

\section{Total Electronic Magnetic Susceptibility - Result and Comparison} \label{SectionTotalSus}

The terms $X^{il}_{\text{dyn}}$ and $X^{il}_{\text{comp}}$ depend on the choice of Wannier functions through their dependence on the $\textbf{k}$-dependent quantities 
\begin{equation}
\label{W}
    \mathcal{W}^a_{nm} = i \sum_{\alpha} \partial_a U_{n\alpha} U^\dag_{\alpha m},
\end{equation}
where $U^{\dag}_{\alpha m} \equiv (U^{\dag})_{\alpha m}$. 
To indicate this we write them as $X^{il}_{\text{dyn}}(\mathcal{W})$ and $X^{il}_{\text{comp}}(\mathcal{W})$, keeping their dependence on the Bloch functions and band energies implicit. Using a generalized inverse effective mass tensor sum rule, and integration by parts over the Brillouin zone, we find that 
\begin{equation}
\label{cancelX}
    X^{il}_\text{dyn}(\mathcal{W})+X^{il}_\text{comp}(\mathcal{W})=0,
\end{equation}
which is shown in detail in section IV of a supplemental material \cite{AlistairAncillaryMagSus}. Therefore the full susceptibility (\ref{dypluscomp}) can be written as \begin{equation}
\label{TotalMagSus}
    \chi^{il} = \chi^{il}_\text{VV} + \chi^{il}_\text{occ} + \chi^{il}_\text{geo} ,
\end{equation}
a main result of this paper. The relation (\ref{cancelX}) holds not only for the $\mathcal{W}^{a}_{nm}$ associated with the exponentially localized Wannier functions that were involved in defining the magnetization, but more generally. In fact, (\ref{cancelX}) holds for \emph{any} unitary transformation $U_{n \alpha}$ that links all the Wannier functions of one type $\alpha$ either to conduction bands exclusively, or to valence bands exclusively; in this situation we have
\begin{equation}
\label{cvbands}
     f_{nm} \mathcal{W}^a_{nm} =0.
\end{equation}

We will see below that the splitting in equation (\ref{TotalMagSus}) of the susceptibility into the three reasonably simple contributions, $\chi^{il}_\text{VV}$, $\chi^{il}_\text{occ}$, and $\chi^{il}_\text{geo}$, is helpful in identifying the ``molecular crystal limit," where the crystal can be considered to consist of molecules at lattice sites between which electrons cannot flow. However, in general these terms are individually gauge dependent. Yet their sum, $\chi^{il}$, is gauge independent.

To see this, consider 
\begin{equation}
\label{simpleBloch}
    \psi_{n \textbf{k}}(\textbf{x}) \rightarrow \psi_{n \textbf{k}}(\textbf{x}) e^{-i\phi_{n} (\textbf{k})},
\end{equation}
and note that if we introduce a unitary matrix $U_{ns}$ associated with this change in Bloch functions, 
\begin{equation}
    U_{ns}=\delta_{ns}e^{-i\phi_n(\textbf{k})},
\end{equation}
where here both $n$ and $s$ indicate Bloch functions, and introduce a $\mathcal{W'}^{a}_{nm}$ analogous to that in (\ref{W}),
\begin{equation}
\label{Wprime}
    \mathcal{W'}^a_{nm} = i \sum_{s} \partial_a U_{ns} U^\dag_{sm},
\end{equation}
we find under the transformation (\ref{simpleBloch}) we have
\begin{equation}
\begin{split}
    &\chi_\text{VV}^{il} \rightarrow \chi_\text{VV}^{il}+X_\text{dyn}^{il}(\mathcal{W'}),
    \\
    &\chi_\text{occ}^{il}+\chi_\text{geo}^{il} \rightarrow \chi_\text{occ}^{il}+\chi_\text{geo}^{il} + X_\text{comp}^{il}(\mathcal{W'})
\end{split}    
\end{equation}
(see section V of \cite{AlistairAncillaryMagSus}). In fact, this holds for more general $U_{ns}(\textbf{k})$ that can mix bands at degenerate points. Since $\mathcal{W'}$ trivially satisfies (\ref{cvbands}), the sum $X_\text{dyn}^{il}(\mathcal{W'})+X_\text{comp}^{il}(\mathcal{W'})$ vanishes, and so $\chi^{il}$ is unchanged.

Of course, the partial integration over the Brillouin zone necessary to establish (\ref{cancelX}) (or the corresponding expression with  $\mathcal{W'}$) involves integration over degeneracy points where the cell-periodic Bloch functions are discontinuous. However, as has been shown in Mahon et al. \cite{ChernPerryJason}, contributions from these points of degeneracy can be written as proportional to the Chern number of the bands, and since we are here concerned with topologically trivial insulators we can expect such contributions to vanish, and gauge invariance is assured.

Despite this gauge invariance, the expression (\ref{TotalMagSus}) for $\chi^{il}$ contains terms with diagonal matrix elements of the Berry connection, and thus it is not convenient for numerical evaluation. We first make comparison to Ogata's work and show that for insulators our expressions agree. Then, starting from those expressions we remove the equal energy elements of the Berry connection. This then allows us to compare to other works in the literature.  

\subsection{Comparison to Ogata}

There is a large body of work on magnetic susceptibilities and their different contributions by Ogata and colleagues (see, e.g., (\cite{OgataMagSusI,OgataMagSusII,OgataMagSusIII,OgataMagSus2017,MatsuuraOgataPeierls}). For low temperature insulators, their expression for the total susceptibility is written as the sum of three terms \cite{OgataMagSus2017},
\begin{equation}
   \label{Ogata_result} \chi^{il}=\chi^{il}_\text{inter}+\chi^{il}_\text{occ}+\chi^{il}_\text{occ2},
\end{equation}
where

\begin{equation}
\label{Ogata_inter}
\begin{split}
    \chi^{il}_\text{inter} = -2\sum_{n\neq m} f_{n} \int_{BZ} \frac{d\textbf{k}}{(2\pi)^3} \text{Re}\Bigg[ \frac{ \mathcal{M}^i_{nm} \Big( \mathcal{M}^l_{nm} \Big)^* }{E_{n\textbf{k}}-E_{m\textbf{k}}}
    \Bigg],  
\end{split}
\end{equation}

\begin{equation}
\label{Ogata_occ2}
\begin{split}
    \chi^{il}_\text{occ2} = -\frac{e}{2\hbar c} \sum_{nn'} f_{n} \int_{BZ} \frac{d\textbf{k}}{(2\pi)^3} \text{Re}\Bigg[ \Omega^i_{nn'}\mathcal{M}^l_{n'n} + \mathcal{M}^i_{nn'} \Omega^l_{n'n} \Bigg], 
\end{split}
\end{equation}
and their term $\chi^{il}_\text{occ}$ is the same as ours (\ref{ChiOcc}). 
However, their interband term $\chi^{il}_\text{inter}$ is written in terms of the non-Hermitian matrix elements $\mathcal{M}^{l}_{nm}$, while our $\chi^{il}_\text{VV}$ is written in terms of the Hermitian matrix elements $M^{l}_{nm}$. And their ``second occupied term" $\chi^{il}_\text{occ2}$ contains a sum over the filled bands $n$, as well as a sum over all bands $n'$ that have the same energy as $n$ over the entire BZ \cite{OgataMagSus2017}, here indicated by the $n$ and $n'$ notation. Note that our $\chi^{il}_\text{geo}$ contains a different integrand and the sums are over all bands. Nonetheless, despite what seem to be significant differences, we show in detail in section VII of the supplemental material that the expressions for the total susceptibilities are equal \cite{AlistairAncillaryMagSus},

\begin{equation}
\begin{split}
 \label{agreement}   \chi^{il}_\text{VV}+\chi^{il}_\text{occ}+\chi^{il}_\text{geo}
    =\chi^{il}_\text{inter} + \chi^{il}_\text{occ} + \chi^{il}_\text{occ2} 
\end{split}
\end{equation}

\subsection{Removal of Equal Energy Berry Connection Matrix Elements} 

We remove the contributions to $\chi^{il}_\text{inter}$, $\chi^{il}_\text{occ}$, and $\chi^{il}_\text{occ2}$ that involve equal energy matrix elements of the Berry connection, and assemble them all in a term we denote by $\underline{\chi}^{il}$. Then indicating by an overset ring the portions of terms that remain when contributions from the equal energy matrix elements of the Berry connection are removed, we have
\begin{equation}
\label{PurefiedSusceptibility}
\begin{split}
    \chi^{il} = \mathring{\chi}^{il}_\text{inter} +  \mathring{\chi}^{il}_\text{occ}  +  \mathring{\chi}^{il}_\text{occ2} + \underline{\chi}^{il} .
\end{split}
\end{equation}
This situation is reminiscent of one we came across in investigating optical activity \cite{PerryOptical,AlistairSipe}. The total optical conductivity tensor we found was gauge-invariant -- i.e., insensitive to an arbitrary complex phase in defining the Bloch functions, as $\chi^{il}$ is here -- and we showed that while individual contributions to the total optical conductivity tensor did depend on equal energy matrix elements of the Berry connection, the total did not. That is, the analog of $\underline{\chi}^{il}$ vanished.  

But here $\underline{\chi}^{il}$ does not vanish. In section VIII of the supplemental material \cite{AlistairAncillaryMagSus} we show that by using integration by parts it can be re-written in a form that does not depend on the equal energy elements of the Berry connection, but instead on a matrix element akin to the Berry curvature, and which is in fact equal to it if there are no degeneracies. We find 
\begin{equation}
\label{chibreve}
    \underline{\chi}^{il} = \frac{1}{2}\mathring{\chi}_\text{occ2:Orb} + \mathring{\chi}_\text{occ2:Spin}.
\end{equation}
Here $\mathring{\chi}_\text{occ2:Orb}$ and $\mathring{\chi}_\text{occ2:Spin}$ result from using (\ref{PureMnn}) for $\mathring{\mathcal{M}}^l_{nn'}$ in place of $\mathcal{M}^l_{nn'}$ in the equation (\ref{Ogata_occ2}) for $\chi^{il}_\text{occ2}$; the ``orbital" contribution  $\mathring{\chi}_\text{occ2:Orb}$ comes from the first term on the right-hand-side of (\ref{PureMnn}) and the ``spin" contribution $\mathring{\chi}_\text{occ2:Spin}$ comes from the second term on the right-hand-side of those equations. That is,
\begin{equation}
\begin{split}
\label{circleOrbSpin}
 \mathring{\chi}^{il}_\text{occ2:Orb,Spin}= 
 \\
 -\frac{e}{2\hbar c} \sum_{nn'} f_{n} \int_{BZ} \frac{d\textbf{k}}{(2\pi)^3} \text{Re}\Bigg[ \mathring{\Omega}^i_{nn'}(\mathring{\mathcal{M}}^l_{n'n})_{\text{Orb,Spin}} \\ + (\mathring{\mathcal{M}}^i_{nn'})_{\text{Orb,Spin}} \mathring{\Omega}^l_{n'n} \Bigg],
 \end{split}
\end{equation}
where 
\begin{equation}
   (\mathring{\mathcal{M}}^l_{nn'})_{\text{Orb}} = \frac{e}{2c} \epsilon^{lab} \sum_{l\neq n} \xi^a_{nl}v^b_{ln'}    
\end{equation}
and
\begin{equation}
    (\mathring{\mathcal{M}}^l_{nn'})_{\text{Spin}} =  \frac{e}{mc} S^l_{nn'}.   
\end{equation}
The modified curl of the non-Abelian Berry connection is defined as \cite{QuantumGeometry}
\begin{equation}
\label{Omegaidentity}   \mathring{\Omega}^i_{nn'}=i\epsilon^{ijk} \sum_{m\neq n} \xi^j_{nm} \xi^k_{mn'},
\end{equation}
 where there is no contribution to the sum in $\mathring{\Omega}^i_{nn'}$ 
at \textbf{k} points where bands $m$ and $n$ or $n'$ are the same or have the same energy, and so the $\mathring{\chi}^{il}_\text{occ2:Orb,Spin}$ are indeed completely independent of those equal energy elements.

The partitioning (\ref{circleOrbSpin}) into orbital and spin contributions is necessary here as they contain different numerical pre-factors in the expression (\ref{chibreve}) for $\underline{\chi}^{il}$. We can now write the total susceptibility as

\begin{equation}
\label{pureTotalSus}
\begin{split}
    \chi^{il} = \mathring{\chi}^{il}_\text{inter} +  \mathring{\chi}^{il}_\text{occ} + \frac{3}{2} \mathring{\chi}_\text{occ2:Orb} + 2 \mathring{\chi}_\text{occ2:Spin},
\end{split}
\end{equation}
where (see the discussion before Eq. (\ref{PurefiedSusceptibility}))
\begin{equation}
   \begin{split}
    \mathring{\chi}^{il}_\text{inter} = -2\sum_{n,m\neq n} f_{n} \int_{BZ} \frac{d\textbf{k}}{(2\pi)^3} \text{Re}\Bigg[ \frac{ \mathring{\mathcal{M}}^i_{nm} \Big( \mathring{\mathcal{M}}^l_{nm} \Big)^* }{E_{n\textbf{k}}-E_{m\textbf{k}}}
    \Bigg],  
\end{split}
\end{equation}
and
\begin{equation}
   \begin{split}
    \mathring{\chi}^{il}_\text{occ} = \frac{e^2}{4\hbar^2 c^2} \epsilon^{iab} \epsilon^{lcd} \sum_{n,m \neq n} f_{n} \int_{BZ} \frac{d\textbf{k}}{(2\pi)^3} \text{Re}\Bigg[ \xi^a_{nm}\xi^c_{mn}  
    \\ \times \Big( \partial_b \partial_d E_{n\textbf{k}} 
    - \frac{\hbar^2}{m} \delta_{bd}\Big) \Bigg].
\end{split} 
\end{equation}
We have thus succeeded in removing all dependence of $\chi^{il}$ on the equal energy elements of the Berry connection.

\subsection{Comparisons to work of Roth, Blount, and Gao et al.}

Work on the magnetic susceptibility was also done by Roth \cite{RothMagSus}, Blount \cite{BlountMagSus}, Hebborn and Sondheimer \cite{HebbornSondheimerEarlyDiamag}, and Gao et al. \cite{GaoGeometricalSus}. Roth's work is notable in that she gave careful consideration to the dependence of the terms in the susceptibility on the phases of the wave functions, noting that diamagnetic and paramagnetic contributions could change but their sum was independent of those phases. Changing the phases alters the ``diagonal position matrix elements," where in modern notation these 
``position matrix elements" are the Berry connection matrix elements we use in this paper. To eliminate any dependence on the diagonal elements, and thus on the Bloch function phases, Roth wrote her expression for the susceptibility with the diagonal elements explicitly removed, in the process of which new terms arose. The expressions of Roth include spin, but are limited to bands that are at most two-fold degenerate over the whole Brillouin zone. When we take this limit in our results, we find agreement with Roth. 

However, Ogata \cite{OgataMagSus2017} pointed out some apparent discrepancies between the results of earlier studies and theirs, and since our expression for an insulator agrees with the low temperature insulator expression of Ogata \cite{OgataMagSus2017}, those apparent discrepancies are also an issue for us.

Our expression -- and that of Ogata et al. (\ref{agreement}) -- seems to differ from earlier work \cite{GaoGeometricalSus,BlountMagSus,RothMagSus} in two respects.  First, in terms in the early work analogous to $\chi^{il}_\text{occ2}$ and $\chi^{il}_\text{geo}$ there is a numerical prefactor different than what both Ogata \cite{OgataMagSus2017} and we find. Second, equal energy elements of the Berry connection are involved in all three contributions $\chi^{il}_\text{inter}$, $\chi^{il}_\text{occ}$, and $\chi^{il}_\text{occ2}$ (and in $\chi^{il}_\text{VV}$ and $\chi^{il}_\text{geo}$), while in earlier work \cite{GaoGeometricalSus,BlountMagSus,RothMagSus} the ``purified" matrix elements  (\ref{PureMnm}, \ref{PureMnn}) that are free of those equal energy terms are used in the expression for the susceptibility; as well, the ``quantum metric" \cite{GaoGeometricalSus} used in earlier work \cite{GaoGeometricalSus,BlountMagSus,RothMagSus} in a term analogous to $\chi^{il}_\text{occ}$ does not contain equal energy elements of the Berry connection. 

Yet we have found that these discrepancies are in fact only apparent. Expression (\ref{pureTotalSus}) is in fact in agreement with earlier work \cite{GaoGeometricalSus,BlountMagSus,RothMagSus}, although Gao et al. \cite{GaoGeometricalSus} and separately Blount \cite{BlountMagSus} only considered the orbital contribution. The details of the comparison are shown in section IX of the supplemental material \cite{AlistairAncillaryMagSus}. Thus our result (\ref{agreement}), which agrees with that of Ogata et al. \cite{OgataMagSusI,OgataMagSusII,OgataMagSusIII,OgataMagSus2017,MatsuuraOgataPeierls} for the limit of low temperature insulators that we consider, is in agreement with earlier work in that limit. This does not seem to have been previously appreciated, and it has apparently been unnoticed that when written in this explicitly gauge-invariant manner the spin and orbital contributions acquire different pre-factors. 

We emphasize that we have not been cavalier with the equal energy elements of the Berry connection, and one evidently should not trivially remove them without care.  And while both expressions (\ref{agreement}) and (\ref{pureTotalSus}) are gauge-invariant, and in agreement with each other, only the second does not involve equal energy elements of the Berry connection and thus is suitable for numerical calculations.  Both expressions, however, involve a sum over all bands at each $\textbf{k}$.

\section{Molecular Crystal Limit}\label{SectionMolecule}

We now consider the ``molecular crystal limit" of our result (\ref{TotalMagSus}, \ref{pureTotalSus}), where we assume that the ``molecule" we associate with each unit cell is far enough away from the other such molecules that electron motion between them can be neglected. In this limit the bands are flat, with $E_{n\textbf{k}}$ independent of $\textbf{k}$.  We then associate a Wannier function type $\alpha$ with each band $n$, $U_{n\alpha}=\delta_{n\alpha}$ and $E_{n\textbf{k}} \rightarrow E_{\alpha}$, identifying the Wannier functions with orbitals of the ``molecules." Taking the limit using the first (\ref{TotalMagSus}) decomposition of the total susceptibility $\chi^{il}$, for the three different contributions we find

\begin{equation}
\begin{split}
     &\chi^{il}_\text{VV} \rightarrow \frac{e^2}{4m^2c^2 \mathcal{V}_{uc}} \sum_{\alpha\beta} \frac{f_{\beta\alpha}}{E_{\alpha}-E_{\beta}} \left(L^l_{\alpha\beta} + 2\bar{S}^l_{\alpha\beta} \right)\left( L^i_{\beta\alpha} + 2\bar{S}^i_{\beta\alpha}  \right),
     \\
     &\chi^{il}_\text{occ} \rightarrow
    -\frac{e^2}{4mc^2\mathcal{V}_{uc}} \epsilon^{iab}\epsilon^{blm} \sum_{\alpha\beta} f_{\alpha} x^a_{\alpha\beta}x^m_{\beta\alpha},
    \\
    &\chi^{il}_\text{geo}\rightarrow 0,
\end{split}     
\end{equation}
where $L_{\alpha\beta}$ are the orbital angular momentum matrix elements between Wannier functions,

\begin{equation}
\label{Lalpha}
    L^i_{\alpha\beta} = \epsilon^{iab}\frac{m}{2} \int d\textbf{x} W^\dag_{\alpha\textbf{0}}(\textbf{x}) \Big( \textbf{x}^a \hat{v}^b(\textbf{x}) \Big) W_{\beta\textbf{0}}(\textbf{x}),
\end{equation}
with $\hat{v}(\textbf{x})$ the term in parentheses on the right hand side of Eq. (\ref{velocity}), and
\begin{equation}
\label{Xalpha}
    x^i_{\alpha\beta}= \int d\textbf{x} W^\dag_{\alpha\textbf{0}}(\textbf{x}) x^i W_{\beta\textbf{0}}(\textbf{x})
\end{equation}
are the position matrix elements. Note that the velocity operator is generalized to include spin-orbit effects (equation (\ref{velocity})). The molecular crystal limit is as one would expect: In this limit $\chi^{il}_{\text{VV}}$ describes the Van Vleck paramagnetism, including the spin magnetic moment contribution, 
\begin{equation}
\label{Salpha}
   \bar{S}^i_{\alpha\beta}= \frac{\hbar}{2}\int d\textbf{x} W^\dag_{\alpha\textbf{0}}(\textbf{x}) \sigma^i W_{\beta\textbf{0}}(\textbf{x}), 
\end{equation}
and the correction to the velocity operator due to SOC, 
and $\chi^{il}_{\text{occ}}$ describes the molecular diamagnetism. In equations (\ref{Lalpha}), (\ref{Xalpha}), (\ref{Salpha}) the Hermitian adjoint of the Wannier functions is used since they are spinors. 

\section{Model Hamiltonian: Gapped Graphene}\label{SectionModelHamiltonian} 

We now turn to a simple model for which the magnetic susceptibility can be calculated explicitly. To fit the scope of the paper we consider a model for an insulator without inversion symmetry, the latter condition so that there is a non-vanishing Berry curvature. A natural choice is a two-band model of a monolayer of hexagonal boron-nitride (gapped graphene). This can be seen as a generalization of a model developed by Ogata \cite{OgataMagSusIII} for the magnetic susceptibility of graphene. As was done there, we consider the $zz$ component of the susceptibility tensor. A principal challenge to any actual calculation with such a model is the fact that only a finite number of bands are taken into account. 
Here this particularly affects the evaluation of the interband contributions to the magnetic susceptibility, and as we see below this affects the trustworthiness of the result. 

\subsection{Tight-Binding Model}

A monolayer of hexagonal boron nitride has two atoms per unit cell. The lattice vectors used in this model are $\textbf{a}_1 = \sqrt{3}a(1,0)$ and $\textbf{a}_2 = \sqrt{3}a(1/2,\sqrt{3}/2)$, where $a = 1.45\mathring{\text{A}}$ is the distance between sites. 
The nearest nitrogen atoms to a boron atom are at displacements $\boldsymbol\delta_{i}$ from the boron atom, where $\boldsymbol\delta_{1} = a(0,-1)$, $\boldsymbol\delta_2 = a(\sqrt{3}/2,1/2)$, and $\boldsymbol\delta_3 = a(-\sqrt{3}/2,1/2)$; see Fig. 1.

\begin{figure}[h]
\includegraphics[width=8cm]{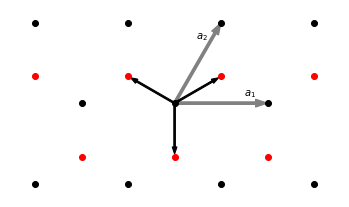}
\caption{Hexagonal boron nitride lattice. The black dots (one at the center) are boron atoms and the red dots (to which the three black arrows point) are nitrogen atoms. The three nearest neighbour lattice vectors (black arrows), and the two primitive lattice vectors (grey arrows) are shown.}
\end{figure}

The cell-periodic part of each Bloch wavefunction is written as a linear combination of orthogonal $p\pi$ orbitals. These orbitals are taken to be
\begin{equation}   
    \phi(\textbf{x}) = \frac{1}{\sqrt{24} (a^*_B)^{5/2}} \sqrt{ \frac{3}{4\pi} } r \cos{\theta} \: e^{-r/2a^*_B},
\end{equation}
where in this formula alone we take the origin to be at a nucleus, $r=|\textbf{x}|$ and $\theta$ is the angle between $\textbf{x}$ and the \textit{z} axis. With $a_B$ the Bohr radius, $a_B^* = a_B/Z_{eff}$ is the effective Bohr radius; $Z_{eff} = 2.4214$ at a boron site and $Z_{eff} = 3.8340$  at a nitrogen site; see Clementi and Raimondi \cite{ClementiRaimondi1963}. Henceforth we use subscripts $A$ and $B$ to refer to boron and nitrogen sites respectively. For the overlap $s=\int \phi_B(\textbf{x}-\textbf{R})\phi_A(\textbf{x}) d\textbf{x}$, where here $\textbf{R}$ is a nearest-neighbor nitrogen site to a boron site at the origin, the orbitals orthogonalized to first order in $s$ are
\begin{equation}
\label{AtomicOrbitalA}
    \Phi_{A}(\textbf{x}-\textbf{R}_A) = \phi_{A}(\textbf{x}-\textbf{R}_A) - \frac{s}{2}\sum_{i=1}^3 \phi_{B}(\textbf{x}-\textbf{R}_A - \boldsymbol\delta_i) 
\end{equation}

\begin{equation}
\label{AtomicOrbitalB}
    \Phi_{B}(\textbf{x}-\textbf{R}_B) = \phi_{B}(\textbf{x}-\textbf{R}_B) - \frac{s}{2}\sum_{i=1}^3 \phi_{A}(\textbf{x}-\textbf{R}_B + \boldsymbol\delta_i) ,
\end{equation}
and the linear combinations of atomic orbitals we use to form the Bloch functions are
\begin{equation}
\label{LCAO}
\begin{split}
    \phi_{A\textbf{k}}(\textbf{x}) = \sqrt{\mathcal{V}_{uc}} \sum_{\textbf{R}_A} e^{-i\textbf{k}\cdot(\textbf{x}-\textbf{R}_A)} \Phi_{A}(\textbf{x}-\textbf{R}_A), 
    \\
    \phi_{B\textbf{k}}(\textbf{x}) = \sqrt{\mathcal{V}_{uc}} \sum_{\textbf{R}_B} e^{-i\textbf{k}\cdot(\textbf{x}-\textbf{R}_B)} \Phi_{B}(\textbf{x}-\textbf{R}_B), 
\end{split}
\end{equation}
where the normalization of these functions is chosen to be consistent with the condition given in the text below equation (\ref{WannierFunc})

The Hamiltonian matrix elements in the basis of the wavefunctions (\ref{LCAO}) are 
\begin{equation}\label{Hamiltonian}
H_{\textbf{k}}  = \begin{pmatrix}
\frac{\Delta}{2}& -t\gamma_{\textbf{k}} \\
-t\gamma_{\textbf{k}}^* & -\frac{\Delta}{2} \\
\end{pmatrix} 
\end{equation}
where $\Delta$ is the gap width, $t$ is the nearest neighbour hopping matrix element, and $\gamma_\textbf{k} = \sum_{i=1}^3 e^{i\textbf{k}\cdot\boldsymbol\delta_{i}}$. The parameters $\Delta$, $t$ are not directly calculated,
but reasonable values are $\Delta=6eV$, $t=3eV$ \cite{GrapheneBasedNanomaterials}, which we adopt here. The dispersion relation is

\begin{equation}
    \epsilon_{\pm} = \pm \frac{1}{2} \sqrt{ 
    \Delta^{2} + 4t^2|\gamma_\textbf{k}|^2
    }
\end{equation}
and the two-band cell-periodic Bloch wavefunctions are
\begin{equation}
\label{cellperiodicfunctions}
    u^{\pm}_\textbf{k} = c^{\pm}_A(\textbf{k}) \phi_{A\textbf{k}} + c_{B}^{\pm}(\textbf{k}) \phi_{B\textbf{k}} 
\end{equation}
where the coefficients are given explicitly, with a specified phase, in Appendix \ref{AppendixhBN}.

To this point two key approximations have been made: (i) we include only nearest-neighbour matrix elements, in the context of a tight-binding model; and (ii) we neglect terms of $O(s^2)$. Note that the second order terms that are dropped are only terms that are explicitly of the form $s^2$. Although the hopping matrix element $t$ is similar to $s$ in that it involves orbitals on neighbouring sites, we do not neglect terms involving the product of $t$ and $s$. To these two approximations we are forced to add a third approximation that (iii) the summations over a complete basis are truncated to a summation over the two bands of the model, since all three of the expressions (\ref{TotalMagSus}, \ref{Ogata_result}, \ref{pureTotalSus}) we have for the susceptibility involve terms that require a summation over all bands.

We focus here on the expressions (\ref{TotalMagSus}, \ref{Ogata_result}) for the susceptibility, since within a tight-binding model the evaluation of the diagonal elements of the Berry connection is not problematic, and we begin with the first of these (\ref{TotalMagSus}) involving the Van Vleck term $\chi^{zz}_{\text{VV}}$, the ``occupied" term $\chi^{zz}_{\text{occ}}$, and the geometric term $\chi^{zz}_{\text{geo}}$.  To evaluate the susceptibility we must calculate three matrix elements (see Appendix \ref{AppendixhBN}),

\begin{equation}
\label{connection}
\begin{split}
    \xi^\mu_{nm} \equiv \frac{i}{\mathcal{V}_{uc}}\int_{\mathcal{V}_{uc}} d\textbf{x} u^\dag_{n\textbf{k}} \frac{\partial u_{m\textbf{k}}}{\partial k_\mu}  
\end{split}
\end{equation}

\begin{equation}
\label{curvature}
    \sum_{s} \xi^{\mu}_{ns} \xi^{\nu}_{sm} = \frac{1}{\mathcal{V}_{uc}}\int_{\mathcal{V}_{uc}} d\textbf{x} \frac{\partial u^\dag_{n\textbf{k}}}{\partial k_{\mu}} \frac{ \partial u_{m\textbf{k}}}{\partial k_\nu} 
\end{equation}

\begin{equation}
\label{angularmomentum}
     \frac{1}{\mathcal{V}_{uc}}\int_{\mathcal{V}_{uc}} d\textbf{x} \frac{\partial u^\dag_{n\textbf{k}}}{\partial k_{\mu}} \frac{\partial H_\textbf{k}}{\partial k_{\nu}} u_{m\textbf{k}} 
\end{equation}
where the first of these is the Berry connection (\ref{Berry connection}), the second appears when evaluating terms that depend on the quantum metric or the Berry curvature, and the third appears when evaluating quantities like the orbital angular momentum. 

The calculation of the susceptibility contributions requires numerical integration 
over the Brillouin zone; we discretize by introducing hexagons of length scaled by $N^{-1}$ 
from that of the BZ area. At $N=250$ there is a 0.4\% error in calculating the BZ area, establishing the scale of numerical error. Results show a clear convergence with increasing $N$ and are evaluated for $N=250$. The values are normalized by
\begin{equation}
    \chi_{0} = \frac{e^2a^2 \mu_{edge}}{8\pi\hbar^2 c^2}
\end{equation}
where $\mu_{edge} = \sqrt{ (3t)^2+(\Delta/2)^2}$ is the band edge energy. That is, in quoting values of the susceptibility components we give them in multiples of $\chi_0$.  

We find $\chi^{zz}_{\text{occ}}=-0.463, \chi^{zz}_{\text{VV}}=0,$ and $\chi^{zz}_\text{geo} = -0.513$. The results indicate a clear diamagnetism of boron nitride, with vanishing Van Vleck term and a significant contribution from both the geometric term and the occupied states term. And although the analytic equivalence (\ref{agreement}) of the two expressions (\ref{TotalMagSus}, \ref{Ogata_result}) follows from sum rules that involve all bands, we find that (\ref{agreement}) holds to very good approximation even with a two-band truncation: For from a direct calculation using two bands we find $\chi^{zz}_{\text{inter}}=0.00808$ and $\chi^{zz}_{\text{occ2}}=-0.519$.

\subsection{Discussion} 
Although the preceding calculation is straightforward, there are a number of curiosities regarding the approximation scheme that we feel are worthy of comment. In the following, we refer to the preceding approximation scheme as ``approximation scheme I." 

Instead of treating only terms explicitly of the form $s^2$ as second order terms to be dropped, one could treat \textit{all} nearest-neighbour matrix elements as first-order quantities, discarding their products. This method was employed earlier by Ogata \cite{OgataMagSusIII}. Under this approximation scheme (``scheme II"), terms linear in $s$ are considered first-order, as they were in scheme I, but as well $t$ is considered a first order quantity, as are all matrix elements of any operator between two orbitals at nearest-neighbour sites. The magnetic susceptibility contributions can be evaluated exactly as before, keeping only first order terms but with this different understanding of what is meant by such terms. Doing so we obtain $\chi^{zz}_{\text{occ}}=-0.466, \chi^{zz}_{\text{VV}}=0,$ and $\chi^{zz}_\text{geo} = -0.515$. These values differ only very slightly from scheme I, and suggest that scheme II is very similar to scheme I.

Nonetheless, there are some issues that arise on closer examination. For instance, in the treatment of graphene by Ogata \cite{OgataMagSusIII}, their equations $(A\cdot 1)$ are first-order quantities with magnitudes of $7a_B^{*2}$ and $1.9a_B^{*2}$, while the magnitude of the corresponding zeroth order quantity, equation $(A\cdot 2)$, is $6a_B^{*2}$; a first order term is larger than the zeroth order term. The same calculation can be done for gapped graphene, with a similar result. As well, using the expression (\ref{Ogata_result}) of $\chi^{il}$ for the gapped graphene model and calculating the contribution $\chi^{zz}_{\text{inter}}$ under scheme II, we obtain a zeroth order term of $-0.0508$ and a first order term of $0.0672$, which again gives a first order term larger in magnitude than the zeroth order term. Note, however, that this is the only contribution where this arises; $\chi^{zz}_{\text{occ2}}$, for example, has a zeroth order term of $-0.640$ and a first order term of $0.113$. Yet these two examples indicate that while scheme II seems \textit{a priori} reasonable, it should be treated with caution. Given that schemes I and II give identical results for $\chi^{zz}$ it seems to appear that the inconsistencies in scheme II tend to cancel out, but there is no reason that this should be universally true. 

One strong motivation for the use of scheme II is that it allows for a complete evaluation of the contribution $\chi^{il}_{\text{inter}}$ in the expression (\ref{Ogata_result}) for the susceptibility (see Appendix \ref{AppendixhBN} and Ogata \cite{OgataMagSusIII}). This still leaves a summation in the contribution $\chi^{il}_\text{occ2}$ that cannot be exactly evaluated, but it does offer some insight into the validity of the two-band approximation. We find $\chi^{zz}_{\text{inter}}=0.205$ under scheme II with a complete evaluation, compared to $\chi^{zz}_{\text{inter}}=0.0164$ under scheme II with a two-band truncation. The difference is significant, and strongly suggests that a two-band truncation is insufficient as an approximation. 

Yet there is one more curiosity to note. The approximation scheme II is implemented analytically. But in numerical evaluations the second order terms are dropped for consistency, due to the product of two terms that are taken up to $O(s)$. Of course, these terms can be calculated and retained, although these are not the only second order terms. In the case of $\chi^{zz}_{\text{inter}}$ these second-order terms have a value of -0.207, the inclusion of which almost entirely cancels the zeroth and first order terms, producing a net small term, as is found from both scheme I and scheme II with a two-band truncation. We emphasize that while this is not a complete second-order evaluation -- it excludes terms arising from second-order terms dropped analytically --- it suggests that the higher order terms cannot be neglected, and since their inclusion brings the value of $\chi^{zz}_{\text{inter}}$ closer to that found from the two-band model, perhaps the two-band truncation may not be entirely unreasonable.

In summary, the major obstacle to calculating the magnetic susceptibility with a tight-binding model is the inability to evaluate the summations over states involving a complete set of bands. The general accuracy of a two-band truncation, as applied here, is unknown. Another strategy (``scheme II"), which both treats overlap integrals and matrix elements between nearest-neighbor sites at the same order, and keeps only first order terms, allows for part of the susceptibility to be calculated without band truncation. But it seems to suffer from certain inconsistencies. These difficulties reveal the limitations of tight-binding models for calculations of the susceptibility, and indicate that full band structure calculations should be employed. For such calculations, the expression (\ref{pureTotalSus}) that avoids the equal energy elements of the Berry connection should be the most useful.

Yet despite the many approximations employed in our h-BN calculation, as outlined above, we compare our result to an experimental measurement of the magnetic susceptibility of h-BN perpendicular to the hexagonal planes. The reported value is a mass susceptibility of $(-0.48\pm0.02)\times10^{-6}$ $cm^3 / g$ \cite{MeasurementhBN}. Using an approximate spacing between successive h-BN layers of $c=3.33\mathring{\text{A}}$ and a density of $\rho = 2.26 g/cm^3$ we obtain a theoretical value of $-0.976\chi_0/(c\rho) = -0.38\times 10^{-6}$ $ cm^3/ g$. This is a little less than 80\% of the measured value.

\section{Conclusions}\label{Conclusions}

In this paper we presented the theoretical expression for the magnetic susceptibility of a topologically trivial insulator in the independent particle approximation, at zero temperature but where both time-reversal and inversion symmetry may be broken. The expression involves a single integral over the Brillouin zone and depends on the band energies and various matrix elements in the cell periodic Bloch function basis. Our result is compared with other theoretical expressions found by very different strategies, including the use of a wave-packet approximation \cite{GaoGeometricalSus}, the application of a Green function framework \cite{FukuyamaEarlyWork}, and the employment of a free energy expansion \cite{BlountMagSus,OgataMagSus2017}. We proved that the apparent differences between the results of the various strategies can be reconciled with the use of certain sum rules such as a generalized effective mass tensor sum rule. Our expression -- and likewise the others \cite{OgataMagSus2017,BlountMagSus,GaoGeometricalSus} that we have shown are equivalent to it -- is gauge-invariant.  

We have also clarified some of the confusion around the appearance of the equal energy elements of the Berry connection in the expression for the magnetic susceptibility. While our first result (\ref{TotalMagSus}) is gauge-invariant, it contains equal energy Berry connection matrix elements, and can be rewritten (\ref{pureTotalSus}) to be independent of these equal energy elements; this is a form more suitable for numerical calculations based on full band structure models. Interestingly, in crystals where at least one of time-reversal or inversion symmetry is broken, this introduces new orbital and spin contributions involving the Berry curvature, but with different prefactors.

Our approach employs a natural decomposition of the magnetization into atomic, itinerant, and spin contributions. This makes the link to the molecular crystal limit, where molecules are identified with lattice sites between which electrons cannot flow, very clear. We recovered the usual Van Vleck term associated with the paramagnetism of the molecules, but including spin effects, together with a contribution associated with the molecular diamagnetism. 

As a sample calculation we considered the susceptibility of hexagonal boron nitride, and examined the consequences of band truncation and the neglect of second order terms in the overlap of orbitals used in tight-binding calculations. In a first calculation (scheme I) the susceptibility of h-BN was determined with all contributions but one evaluated using a two-band truncation, while in a second (scheme II) the interband paramagnetic contribution to the susceptibility can be evaluated ``exactly" upon neglect of second order in the orbital overlap terms.  The second calculation yields a result about 12 times larger than that of the first scheme, putting some doubt on the validity of the two-band truncation.  However, in the second scheme the first order orbital overlap terms in one contribution are larger than the zeroth order terms, and the second order terms that had to be discarded to make the complete evaluation are larger still.  While this comparison can only be made for one contribution, it also raises doubts about the validity of approximate treatments of the orbital overlaps.  Both these issues suggest that full electronic structure calculations would be in order, and necessary to obtain trustworthy results.

Also looking to the future, a general benefit of our approach is that it can be extended to investigate the response of a material to finite frequency and spatially varying fields, an extension not easily implemented using other strategies. When this is done, higher order multipole moments make contributions to the induced charge current densities.  For example, when optical activity was treated with this approach it required the identification of the electric quadrupole response to the electric field, the electric dipole response to symmetrized derivatives of the electric field and the magnetic field, and the magnetic dipole response to an electric field \cite{PerryOptical}. These moments combine to form a response of the charge-current density to applied fields that is gauge-invariant.  The moments and their responses that are responsible for a generalization of the magnetic susceptibility are easily identified, and we plan to present the associated response calculation in a future publication. 

\section{Acknowledgements} 

We thank Jason Kattan for helpful comments and discussions. This work was supported by
the Natural Sciences and Engineering Research Council of
Canada (NSERC).  A.H.D.
acknowledges a PGS-D scholarship from NSERC. 

\appendix 

\begin{widetext}
\section{Relevant Formulae for Derivation}\label{Appendix:relegatedexpressions} 

In this appendix we present 
expressions that are important for the derivation of the 
various contributions to the magnetic susceptibility. 
For the derivation of these expressions themselves, see Duff and Sipe \cite{AlistairSipe}.

The velocity charge current $\textbf{j}^{\mathfrak{p}}_\textbf{R}(\textbf{x},t)$ 
in equation (\ref{atomicmag}) is

\begin{equation}
    \textbf{j}^{\mathfrak{p}}_\textbf{R}(\textbf{x},t) = \sum_{\alpha,\beta, \textbf{R}',\textbf{R}''} \textbf{j}^{\mathfrak{p}}_{\beta \textbf{R}';\alpha \textbf{R}''}(\textbf{x},\textbf{R};t) \eta_{\alpha \textbf{R}'';\beta \textbf{R}'}(t),
\end{equation}
with

\begin{equation}
\begin{split}
    \textbf{j}^{\mathfrak{p}}_{\beta \textbf{R}';\alpha \textbf{R}''}(\textbf{x},\textbf{R};t) =& \Bigg(\frac{1}{2}\delta_{\textbf{R}\textbf{R}''}e^{i\Delta(\textbf{R}',\textbf{x},\textbf{R}'';t)}\chi^*_{\beta \textbf{R}'; \textsf{j}} (\textbf{x},t) \left(\textbf{J}^\mathfrak{p}_\textsf{ji}(\textbf{x},\boldsymbol{\mathfrak{p}}(\textbf{x},\textbf{R};t))\chi_{\alpha\textbf{R}'';\textsf{i}}(\textbf{x},t) \right)
    \\
    &+\frac{1}{2}\delta_{\textbf{R}\textbf{R}''}\left(\textbf{J}^\mathfrak{p}_\textsf{ji}(\textbf{x},\boldsymbol{\mathfrak{p}}^*(\textbf{x},\textbf{R};t))e^{i\Delta(\textbf{R}',\textbf{x},\textbf{R}'';t)}\chi^*_{\beta \textbf{R}' \textsf{j}}(\textbf{x},t)\right) \chi_{\alpha\textbf{R}''\textsf{i}}(\textbf{x},t) 
    \\
    &+\frac{1}{2}\delta_{\textbf{R}\textbf{R}'}\chi^*_{\beta \textbf{R}' \textsf{j}}(\textbf{x},t) \left(\textbf{J}^\mathfrak{p}_\textsf{ji}(\textbf{x},\boldsymbol{\mathfrak{p}}(\textbf{x},\textbf{R};t)) e^{i\Delta(\textbf{R}',\textbf{x},\textbf{R}'';t)}\chi_{\alpha\textbf{R}''\textsf{i}}(\textbf{x},t) \right) 
    \\
    &+\frac{1}{2}\delta_{\textbf{R}\textbf{R}'}\left(\textbf{J}^\mathfrak{p}_\textsf{ji}(\textbf{x},\boldsymbol{\mathfrak{p}}^*(\textbf{x},\textbf{R};t))\chi^*_{\beta \textbf{R}' \textsf{j}}(\textbf{x},t)\right)e^{i\Delta(\textbf{R}',\textbf{x},\textbf{R}'';t)} \chi_{\alpha\textbf{R}''\textsf{i}}(\textbf{x},t) 
    \Bigg).
\end{split}
\end{equation}
where

\begin{equation}
\label{Jp}
\begin{split}
    \textbf{J}^{\mathfrak{p}}_\textsf{ij}(\textbf{x},\mathfrak{p}(\textbf{x},\textbf{y};t)) = \frac{e}{2m}\left(\boldsymbol{\mathfrak{p}}(\textbf{x})- \frac{e}{c}\Omega_\textbf{y}(\textbf{x},t)\right)\delta_\textsf{ij} 
    + \epsilon_{abc}\hat{e}^c \frac{e\hbar}{8m^2c^2}\sigma_\textsf{ij}^a\frac{ \partial  \text{V}(\textbf{x})}{\partial x_b}.
\end{split}
\end{equation}
Here sans-serif indices 
indicate spinor components.
The vector potential has been replaced by a relator-dependent quantity
\begin{equation}
    \Omega_\textbf{y}^k(\textbf{x},t) \equiv \int \alpha^{lk}(\textbf{w};\textbf{x},\textbf{y}) B^l(\textbf{w},t) d\textbf{w},
\end{equation}

The itinerant charge current $\tilde{\textbf{j}}_\textbf{R}(\textbf{x},t)$ is defined as

\begin{equation}
    \tilde{\textbf{j}}_\textbf{R}(\textbf{x},t) = \sum_{\alpha,\beta,\textbf{R}',\textbf{R}''} \tilde{\textbf{j}}_{\beta\textbf{R}';\alpha\textbf{R}''}(\textbf{x},\textbf{R};t) \eta_{\alpha\textbf{R}'';\beta\textbf{R}'}(t),
\end{equation}
with
\begin{equation}
    \tilde{\textbf{j}}_{\beta\textbf{R}';\alpha\textbf{R}''}(\textbf{x},\textbf{R};t) = \frac{1}{2}(\delta_{\textbf{R}\textbf{R}''} + \delta_{\textbf{R}\textbf{R}'}) \tilde{\textbf{j}}_{\beta\textbf{R}';\alpha\textbf{R}''}(\textbf{x},t),
\end{equation}
and
\begin{equation}
\begin{split}
    \tilde{\textbf{j}}_{\beta \textbf{R}''';\alpha \textbf{R}''}(\textbf{x},t) =& -\sum_\textbf{R} \int s(\textbf{x};\textbf{y},\textbf{R})\Gamma_\textbf{R}^{\alpha\textbf{R}'';\beta\textbf{R}'''}(\textbf{y},t) d\textbf{y}
    - \frac{1}{2}\sum_{\textbf{R},\textbf{R}'} s(\textbf{x};\textbf{R},\textbf{R}') \zeta_{\textbf{R}\textbf{R}'}^{\alpha\textbf{R}'';\beta\textbf{R}'''}(t),
\end{split}
\end{equation}
where
\begin{equation}
\begin{split}
    \zeta_{\textbf{R}\textbf{R}'}^{\alpha\textbf{R}'';\beta\textbf{R}'''}(t) = \frac{e}{i\hbar}\Big(
    \delta_{\textbf{R}'''\textbf{R}}\delta_{\textbf{R}''\textbf{R}'}\bar{H}_{\beta\textbf{R};\alpha\textbf{R}'}(t) 
    - \delta_{\textbf{R}''\textbf{R}}\delta_{\textbf{R}'''\textbf{R}'}\bar{H}_{\beta\textbf{R}';\alpha\textbf{R}}(t)
    \Big),
\end{split}
\end{equation}
and 
\begin{equation}
    \begin{split}
        \Gamma_\textbf{R}^{\alpha\textbf{R}'';\beta\textbf{R}'}(\textbf{x},t) = \nabla \cdot \textbf{j}^{\mathfrak{p}}_{\beta \textbf{R}';\alpha \textbf{R}''}(\textbf{x},\textbf{R};t)
        + \frac{\partial \rho_{\beta\textbf{R}';\alpha\textbf{R}''}(\textbf{x},\textbf{R};t)}{\partial t}
        + \frac{1}{i\hbar} \sum_{\mu,\nu,\textbf{R}_1,\textbf{R}_2}\rho_{\nu\textbf{R}_2;\mu\textbf{R}_1}(\textbf{x},\textbf{R};t)\mathfrak{F}^{\alpha\textbf{R}'';\beta\textbf{R}'}_{\mu\textbf{R}_1;\nu \textbf{R}_2}(t),
    \end{split}
\end{equation}
with
\begin{equation}
\begin{split}
    \mathfrak{F}^{\alpha\textbf{R}'';\beta\textbf{R}'}_{\mu\textbf{R}_1;\nu\textbf{R}_2}(t) = \delta_{\beta\nu}\delta_{\textbf{R}_2\textbf{R}'}e^{i\Delta(\textbf{R}_1,\textbf{R}'',\textbf{R}_2;t)}\bar{H}_{\mu\textbf{R}_1;\alpha\textbf{R}''}(t)
    -\delta_{\alpha\mu} \delta_{\textbf{R}''\textbf{R}_1}e^{i\Delta(\textbf{R}_1,\textbf{R}',\textbf{R}_2;t)}\bar{H}_{\beta\textbf{R}';\nu\textbf{R}_2}(t) \\
    - e\delta_{\beta\nu}\delta_{\alpha\mu}\delta_{\textbf{R}_2\textbf{R}'}\delta_{\textbf{R}_1\textbf{R}''}\Omega^0_{\textbf{R}_2}(\textbf{R}_1;t),
\end{split}
\end{equation}
and

\begin{equation}
\label{HMatrix}
\begin{split}
    \bar{H}_{\alpha\textbf{R};\lambda\textbf{R}''}(t) =
    \frac{1}{2}  \int d\textbf{x}& \chi^*_{\alpha\textbf{R};\textsf{i}}(\textbf{x},t)e^{i\Delta(\textbf{R},\textbf{x},\textbf{R}'';t)}
    (\mathcal{K}_\textsf{ik}(\textbf{x},\textbf{R}'';t)\chi_{\lambda \textbf{R}'';\textsf{k}}(\textbf{x},t))
    \\
    + \frac{1}{2} \int d\textbf{x}& (\mathcal{K}_\textsf{ki}(\textbf{x},\textbf{R};t)\chi_{\alpha\textbf{R};\textsf{i}}(\textbf{x},t))^*e^{i\Delta(\textbf{R},\textbf{x},\textbf{R}'';t)}\chi_{\lambda \textbf{R}'';\textsf{k}}(\textbf{x},t)
    \\
    -\frac{i\hbar}{2}\int d\textbf{x}& e^{i\Delta(\textbf{R},\textbf{x},\textbf{R}'';t)} \Bigg( \chi^*_{\alpha \textbf{R};\textsf{i}}(\textbf{x},t) \frac{\partial \chi_{\lambda \textbf{R}'';\textsf{i}}(\textbf{x},t)}{\partial t} - 
    \frac{\partial \chi^*_{\alpha \textbf{R};\textsf{i}}(\textbf{x},t)}{\partial t}\chi_{\lambda \textbf{R}'';\textsf{i}}(\textbf{x},t)
    \\
    &-\frac{ie}{\hbar}\big(\Omega^0_{\textbf{R}''}(\textbf{x},t)+\Omega^0_\textbf{R}(\textbf{x},t)\big)\chi^*_{\alpha\textbf{R};\textsf{i}}(\textbf{x},t) \chi_{\lambda \textbf{R}'';\textsf{i}}(\textbf{x},t)
    \Bigg),
\end{split}
\end{equation}

\begin{equation}
    \Omega^0_\textbf{y}(\textbf{x},t) \equiv \int s^i(\textbf{w};\textbf{x},\textbf{y}) E^i(\textbf{w},t) d\textbf{w},
\end{equation}
and the operator $\mathcal{K}_\textsf{ij}(\textbf{x},\textbf{y};t)$ is defined as
\begin{equation}
\begin{split}
    \mathcal{K}_\textsf{ij}(\textbf{x},\textbf{y};t) = \frac{(\mathfrak{p}(\textbf{x})-\frac{e}{c}\Omega_\textbf{y}(\textbf{x},t))^2}{2m} \delta_\textsf{ij} + \text{V}(\textbf{x}) \delta_\textsf{ij} - \frac{e\hbar}{2mc} \boldsymbol\sigma_\textsf{ij} \cdot \textbf{B}(\textbf{x},t) - \frac{e\hbar}{2mc} \boldsymbol\sigma_\textsf{ij} \cdot \textbf{B}_\text{static}(\textbf{x})
    \\
    +\frac{\hbar}{4m^2 c^2} \boldsymbol \sigma_\textsf{ij} \cdot \Big( \nabla\text{V}(\textbf{x}) \times ( \mathfrak{p}(\textbf{x}) - \frac{e}{c}\boldsymbol{\Omega}_\textbf{y}(\textbf{x},t) ) \Big).
\end{split}
\end{equation}


\section{Atomic, Itinerant, and Spin Susceptibility}\label{Appendix:SusceptibilityCombinations}

While in the body of this paper we have divided the magnetic susceptibility into the Van Vleck, occupied, and geometric terms, the 
more natural decomposition of the susceptibility in our approach is into 
atomic, itinerant, and spin magnetizations and their respective dynamical and compositional contributions. 
For a derivation of the following 
expressions see section III of Duff and Sipe \cite{AlistairAncillaryMagSus}. 

Beginning with the dynamical contributions, the atomic magnetization dynamical contribution is

\begin{equation}
\begin{split}
    \bar{M}^{i}_\text{dyn} = \frac{e}{4 c} \epsilon^{icd}  B^l \sum_{mns} f_{nm} \int_{BZ} \frac{d\textbf{k}}{(2\pi)^3} 
    \frac{ \Big(v^d_{ns}\xi^c_{sm}+\xi^c_{ns}v^d_{sm} \Big)M^l_{mn} }{\Delta_{mn}(\textbf{k})} 
    \\
    + \frac{e}{4c} \epsilon^{icd} B^l \sum_{mns} f_{nm} \int_{BZ} \frac{d\textbf{k}}{(2\pi)^3} \frac{ \Big( \mathcal{W}^c_{ns} v^d_{sm} + v^d_{ns}\mathcal{W}^c_{sm} \Big) M^l_{mn} }{\Delta_{mn}(\textbf{k})}
    \\
    + \frac{ie^2}{16\hbar c^2} \epsilon^{icd} \epsilon^{lab} B^l \sum_{mnls} f_{nm} \int_{BZ} \frac{d\textbf{k}}{(2\pi)^3} \xi^b_{mn} \Bigg( \Big((\xi^c_{nl} + \mathcal{W}^c_{nl})v^d_{ls} + v^d_{nl}(\xi^c_{ls}+\mathcal{W}^c_{ls}) \Big)\mathcal{W}^a_{sm}
    \\
    + \mathcal{W}^a_{ns}\Big( (\xi^c_{sl} + \mathcal{W}^c_{sl}) v^d_{lm} + v^d_{sl} (\xi^c_{lm} + \mathcal{W}^c_{lm}) \Big) \Bigg),
\end{split}
\end{equation}
where
\begin{equation}
    \mathcal{W}^a_{ns} = i \sum_{\alpha} \partial_a U_{n\alpha} U^\dag_{\alpha s}.
\end{equation}
The itinerant dynamical contribution is
\begin{equation}
\begin{split}
    \tilde{M}^{i}_\text{dyn} = -\frac{e}{4\hbar c} \epsilon^{icd} B^l \sum_{mns} f_{nm} \int_{BZ} \frac{d\textbf{k}}{(2\pi)^3} \text{Re}\Bigg[
    \frac{ \xi^d_{nm} \partial_c (E_{m\textbf{k}}+E_{n\textbf{k}}) M^l_{mn} }{ \Delta_{mn}(\textbf{k}) }
    \Bigg] 
    \\
    -\frac{ie}{4\hbar c} \epsilon^{icd} B^l \sum_{mns} f_{nm} \int_{BZ} \frac{d\textbf{k}}{(2\pi)^3} \frac{
    \Big((E_{l\textbf{k}}-E_{n\textbf{k}}) \mathcal{W}^c_{nl} (\xi^d_{lm} + \mathcal{W}^d_{lm}) + (E_{m\textbf{k}}-E_{l\textbf{k}}) (\xi^d_{nl}+\mathcal{W}^d_{nl}) \mathcal{W}^c_{lm} \Big)
    M^l_{mn} }{\Delta_{mn}(\textbf{k})}
    \\
    -\frac{e^2}{8\hbar^2 c^2} \epsilon^{icd} \epsilon^{lab} B^l \sum_{mn} f_{nm} \int_{BZ} \frac{d\textbf{k}}{(2\pi)^3} \text{Re}\Bigg[ 
    (E_{n\textbf{k}}-E_{l\textbf{k}}) \mathcal{W}^c_{nl} (\xi^d_{ls} + \mathcal{W}^d_{ls})\mathcal{W}^a_{sm} \xi^b_{mn} 
    \\ 
    + (E_{l\textbf{k}}-E_{s\textbf{k}}) (\xi^d_{nl}+\mathcal{W}^d_{nl})\mathcal{W}^c_{ls} \mathcal{W}^a_{sm} \xi^b_{mn} 
    + i \partial_c (E_{n\textbf{k}}+E_{l\textbf{k}}) (\xi^d_{nl}+\mathcal{W}^d_{nl}) \mathcal{W}^a_{lm} \xi^b_{mn}
    \Bigg]
    ,
\end{split}
\end{equation}
and lastly the spin dynamical contribution is
\begin{equation}
\begin{split}
    \breve{M}^{i}_{dyn} = & 
    \frac{e\hbar}{2mc}  B^l \sum_{mn} f_{nm} \int_{BZ} \frac{d\textbf{k}}{(2\pi)^3}\frac{S^i_{nm} M^l_{mn}}{\Delta_{mn}(\textbf{k})}
    \\
    + &\frac{ie^2}{4\hbar mc^2} \epsilon^{lab} B^l \sum_{mn} f_{nm} \int_{BZ} \frac{d\textbf{k}}{(2\pi)^3} \Big( S^i_{ns}\mathcal{W}^a_{sm} + \mathcal{W}^a_{ns}S^i_{sm} \Big)\xi^b_{mn}.
\end{split}
\end{equation}
The combination of the three dynamical contributions is
\begin{equation}
\begin{split}
    \bar{M}^{i}_\text{dyn}+ \tilde{M}^{i}_\text{dyn}+ \breve{M}^{i}_\text{dyn} 
    \\
    = B^l \sum_{mn} f_{nm} \int_{BZ}\frac{d\textbf{k}}{(2\pi)^3}\Bigg( \frac{e}{4c}\epsilon^{icd}\Big(v^d_{ns}\xi^c_{sm}+\xi^c_{ns}v^d_{sm} -\frac{1}{\hbar}\xi^d_{nm} \partial_c(E_{n\textbf{k}}+E_{m\textbf{k}}) \Big) + \frac{e\hbar}{2mc} S^i_{nm} \Bigg) \frac{ M^l_{mn}}{\Delta_{mn}(\textbf{k})}
    \\
    -B^l\frac{ie}{4\hbar c}  \sum_{mn} f_{nm} \int_{BZ} \frac{d\textbf{k}}{(2\pi)^3} \epsilon^{icd} \Bigg( \mathcal{W}^c_{nl}\xi^d_{lm} + \xi^d_{nl} \mathcal{W}^c_{lm} \Bigg) M^l_{mn} 
    \\
    +B^l\frac{ie}{4\hbar c} \epsilon^{lab} \sum_{mn} f_{nm} \int_{BZ} \frac{d\textbf{k}}{(2\pi)^3} M^i_{nm}\Big( \mathcal{W}^a_{ms}\xi^b_{sn} + \xi^b_{ms}\mathcal{W}^a_{sn}\Big) 
    \\
    -B^l\epsilon^{icd}\epsilon^{lab} \frac{e^2}{8\hbar^2 c^2} \sum_{mn} f_{nm} \int_{BZ} \frac{d\textbf{k}}{(2\pi)^3} \text{Re}\Bigg[ (E_{n\textbf{k}}-E_{m\textbf{k}}) \mathcal{W}^c_{nl} \xi^d_{lm} \mathcal{W}^a_{ms}\xi^b_{sn} + (E_{n\textbf{k}}-E_{m\textbf{k}}) \mathcal{W}^c_{nl} \xi^d_{lm} \xi^b_{ms} \mathcal{W}^a_{sn} \Bigg],
    \\
    = B^l \sum_{mn} f_{nm} \int_{BZ} \frac{d\textbf{k}}{(2\pi)^3} \frac{ M^i_{nm} M^l_{mn} }{\Delta_{mn}(\textbf{k})} + X^{il}_\text{dyn} B^l
    \\
    = \Big(\chi^{il}_\text{VV} + X^{il}_\text{dyn} \Big) B^l.
\end{split}
\end{equation}

The compositional contributions are more compactly written in the Wannier function basis. To convert to the Bloch function basis one has to implement the following relationships
\begin{equation}
\begin{split}
    &\tilde{v}^a_{\alpha\beta} = \sum_{mn} U^\dag_{\alpha n} v^a_{nm} U_{m\beta},
    \\
    &\tilde{\xi}^a_{\alpha\beta} = \sum_{mn} U^\dag_{\alpha n} (\xi^a_{nm} + \mathcal{W}^a_{nm}) U_{m\beta},
    \\
    & \tilde{S}^a_{\alpha\beta} = \sum_{mn} U^\dag_{\alpha n} S^a_{nm} U_{m\beta} .
\end{split}
\end{equation}
The atomic magnetization compositional contribution can be written as 
\begin{equation}
\begin{split}
    \bar{M}^{i}_\text{comp} = -\frac{e^2}{8\hbar c^2}  \epsilon^{icd} B^l \epsilon^{lab} \sum_{\alpha\beta} f_{\alpha} \text{Re}\Bigg[ \int_{BZ} \frac{d\textbf{k}}{(2\pi)^3} \Big( \tilde{\xi}^c_{\alpha\gamma}\tilde{v}^d_{\gamma\beta} + \tilde{v}^d_{\alpha\gamma} \tilde{\xi}^c_{\gamma\beta}
    -i\partial_c \tilde{v}^d_{\alpha\beta} 
    \Big) \partial_a \tilde{\xi}^b_{\beta\alpha} \Bigg]
    \\
    -\epsilon^{icd}\epsilon^{dab} B^a \frac{e^2}{4mc^2} \sum_{\alpha\beta} f_{\alpha} \int_{BZ} \frac{d\textbf{k}}{(2\pi)^3} \tilde{\xi}^c_{\alpha\beta} \tilde{\xi}^b_{\beta\alpha}, 
\end{split}
\end{equation}
the 
itinerant magnetization compositional contribution is
\begin{equation}
\begin{split}
    \tilde{M}^{i}_\text{comp} = \frac{e^2}{4\hbar^2 c^2} B^l \epsilon^{iab} \epsilon^{lcd} \sum_{\alpha\beta\gamma} f_{\alpha} \text{Re}\Bigg[ 
    \int_{BZ} \frac{d\textbf{k}}{(2\pi)^3} \tilde{\xi}^d_{\alpha\gamma} \tilde{\xi}^b_{\gamma\beta} \partial_a \partial_c \Big( U^\dag_{\beta n} E_{n\textbf{k}} U_{n\alpha} \Big) 
    \Bigg]
    \\
    + \frac{e^2}{8\hbar^2c^2} \epsilon^{iab} \epsilon^{lcd} B^l \sum f_{\alpha} \text{Re}\Bigg[ \int_{BZ} \frac{d\textbf{k}}{(2\pi)^3}  \Big( \partial_c \tilde{\xi}^d_{\alpha\beta} \tilde{\xi}^b_{\beta \mu} + \tilde{\xi}^b_{\alpha\beta} \partial_c \tilde{\xi}^d_{\beta \mu} \Big) \partial_a \Big( U^\dag_{\mu n} E_{n\textbf{k}} U_{n\alpha} \Big) \Bigg] 
    \\
    -\frac{e^2}{8\hbar c^2} \epsilon^{iab} \epsilon^{lcd} B^l \sum_{\alpha} f_{\alpha} \int_{BZ} \frac{d\textbf{k}}{(2\pi)^3} \text{Re}\Bigg[
    \partial_a \tilde{\xi}^b_{\alpha\mu} \Bigg( \tilde{\xi}^c_{\mu\gamma} \tilde{v}^d_{\gamma\alpha} + \tilde{v}^d_{\mu\gamma} \tilde{\xi}^c_{\gamma\alpha} - \frac{1}{\hbar} \Big( \tilde{\xi}^d_{\mu\beta} \partial_c \Big( U^\dag_{\beta n} E_{n\textbf{k}} U_{n\alpha}\Big) + \partial_c \Big(U^\dag_{\mu n} E_{n\textbf{k}} U_{n\beta} \Big) \tilde{\xi}^d_{\beta\alpha} \Big)
    \Bigg]
    \\
    -\frac{e^2}{2\hbar m c^2} \epsilon^{iab} B^l \sum f_{\alpha} \text{Re}\Bigg[ \int_{BZ} \frac{d\textbf{k}}{(2\pi)^3} \partial_a \tilde{\xi}^b_{\alpha\mu} \tilde{S}^l_{\mu\alpha} \Bigg],
\end{split}
\end{equation}
and the spin magnetization compositional contribution is
\begin{equation}
    \breve{M}^{i}_\text{comp}= -\frac{e^2}{2m\hbar c^2} \epsilon^{lab} B^l \sum f_{\alpha} \text{Re}\Bigg[  \int_{BZ} \frac{d\textbf{k}}{(2\pi)^3} \tilde{S}^i_{\alpha\beta} \partial_a \tilde{\xi}^b_{\beta\alpha}   \Bigg].
\end{equation}
In the following the conversion to the Bloch function basis has been done, but the details are omitted as they are very involved; they can be found in section III of an 
ancillary document \cite{AlistairAncillaryMagSus}. Noting that 
the curl of the Berry connection 
is the non-Abelian (and non gauge-covariant) Berry curvature $\Omega^{i}_{nm}$ (\ref{nonAbelian}), we can combine all the terms with the `i'th' or `l'th' component of $\boldsymbol\Omega_{nm}$:

\begin{equation}
\begin{split}
    \bar{M}^{i}_\text{comp} + \tilde{M}^{i}_\text{comp} + \breve{M}^{i}_\text{comp} 
    \\
    =\frac{e}{2\hbar c}  B^l \sum_{mns} f_{n} \int_{BZ} \frac{d\textbf{k}}{(2\pi)^3} \text{Re}\Bigg[
    \Omega^{i}_{nm} \Bigg(\frac{e}{4c} \epsilon^{lcd} \Big(v^c_{ms}\xi^d_{sn} +\xi^d_{ms}v^c_{sn} + \frac{1}{\hbar} \xi^d_{mn} \partial_c (E_{n\textbf{k}}+E_{m\textbf{k}})\Big) - \frac{e}{mc}S^l_{mn}\Bigg)
    \Bigg]
    \\
    -\frac{e}{2\hbar c} B^l \sum_{mns} f_{n} \int_{BZ} \frac{d\textbf{k}}{(2\pi)^3} \text{Re}\Bigg[ 
    \Omega^l_{mn} \Bigg(\frac{e}{4c}\epsilon^{iab}\Big( \xi^a_{ns}v^b_{sm}+v^b_{ns}\xi^a_{sm} - \frac{2}{\hbar}\partial_a E_{n\textbf{k}} \xi^b_{nm} + i\partial_a v^b_{nm}\Big)
    + \frac{e}{m c} S^i_{nm}
    \Bigg)
    \Bigg]
    \\
    -\frac{e^2}{4mc^2} B^l \sum_{mn} f_{n} \int_{BZ} \frac{d\textbf{k}}{(2\pi)^3} \text{Re}\Bigg[
    \epsilon^{iab}\epsilon^{lcd}\Big(\delta_{bd} \xi^a_{nm}\xi^c_{mn} - \frac{m}{\hbar^2}  \xi^b_{nm}\xi^d_{mn} \partial_a \partial_c E_{n\textbf{k}}\Big)
    \Bigg]
    \\
    + X^{il}_\text{comp},
\end{split}
\end{equation}
where 
$X^{il}_\text{comp}$ depends on the $\mathcal{W}^a_{nm}$
; it thus tracks the dependence of the total expression on the Wannier function basis, and also characterizes how that total expression depends on the Bloch bundle gauge freedom.
Inside the brackets of the first line we see the spontaneous magnetization matrix element; see equation (\ref{SpontMagEq}). In the second line 
the appearance of the spontaneous magnetization matrix elements is not immediately apparent; we must take the derivative of the velocity matrix element to find
\begin{equation}
\begin{split}
    \Bigg(\frac{e}{4c}\epsilon^{iab}\Big( \xi^a_{ns}v^b_{sm}+v^b_{ns}\xi^a_{sm} - \frac{2}{\hbar}\partial_a E_{m\textbf{k}} \xi^b_{nm} - \frac{1}{\hbar}\partial_a \Big((E_{n\textbf{k}}-E_{m\textbf{k}})\xi^b_{nm} \Big)\Big)
    + \frac{e}{m c} S^i_{nm}
    \Bigg)
    \\
    = \Bigg(\frac{e}{4c}\epsilon^{iab}\Big( \xi^a_{ns}v^b_{sm}+v^b_{ns}\xi^a_{sm} - \frac{1}{\hbar}\partial_a (E_{n\textbf{k}} + E_{m\textbf{k}})\xi^b_{nm}\Big) + \frac{e}{mc} S^i_{nm} -\epsilon^{iab} \frac{e}{4\hbar c} (E_{n\textbf{k}}-E_{m\textbf{k}}) \partial_a \xi^b_{nm}\Bigg)
    \\
    = M^i_{nm} - \frac{e}{4\hbar c} (E_{n\textbf{k}}-E_{m\textbf{k}})\Omega^i_{nm}.
\end{split}
\end{equation}

With this identification 
we can write the total compositional response as
\begin{equation}
\begin{split}
    \bar{M}^{i,(B,I)} + &\tilde{M}^{i,(B,I)} + \breve{M}^{i,(B,I)} 
    \\
    =& -\frac{e}{2\hbar c} B^l \sum_{mns} f_{n} \int_{BZ} \frac{d\textbf{k}}{(2\pi)^3} \text{Re}\Bigg[
    \Omega^i_{nm} M^l_{mn} + M^i_{nm}\Omega^l_{mn}
    \Bigg]
    \\
    &-\frac{e^2}{8\hbar^2c^2} B^l \sum_{mn} f_{n} \int_{BZ} \frac{d\textbf{k}}{(2\pi)^3} \text{Re}\Bigg[
    (E_{n\textbf{k}}-E_{m\textbf{k}})\Omega^i_{nm}\Omega^l_{mn}
    \Bigg] 
    \\
    &-\frac{e^2}{4mc^2} \epsilon^{iab}\epsilon^{lcd} B^l \sum_{mn} f_{n} \int_{BZ} \frac{d\textbf{k}}{(2\pi)^3} \Bigg[
    \delta_{bd} \xi^a_{mn}\xi^c_{nm} - \frac{m}{\hbar^2}  \xi^b_{mn}\xi^d_{nm} \partial_a \partial_c E_{n\textbf{k}}
    \Bigg]
    + X^{il}_\text{comp}.
    \\
    = &\Big(\chi^{il}_\text{geo}+\chi^{il}_\text{occ} + X^{il}_\text{comp} \Big) B^l.
\end{split}
\end{equation}

\end{widetext}

\section{Details of h-BN model calculations}\label{AppendixhBN}

The coefficients for the wavefunctions are related by 
\begin{equation}
\label{ABcoefficients}
    c^{\pm}_{A}/c^{\pm}_{B} = \alpha^{\pm}(\textbf{k}), 
\end{equation}
where
\begin{equation}
    \alpha^{\pm}(\textbf{k}) = \frac{t \gamma_\textbf{k}}{\frac{\Delta}{2} - \epsilon^{\pm}}.
\end{equation}
The relative phase of the coefficients is the same as that in graphene, which experiences a discontinuity at the Dirac points. However, in boron nitride the coefficients also have different magnitudes, and at points of phase discontinuity (the Dirac points of graphene), one of the coefficients goes to zero: $c_B$ in the conduction band, $c_{A}$ in the valence band. The coefficients themselves are well defined if the complex phase is put entirely on the coefficient that vanishes at the point of phase singularity. Thus the coefficients are given by 
\begin{equation}
\label{Acoefficients}
    c^{+}_{A} = \frac{1}{\sqrt{1 + |1/\alpha^+|^2}} \hspace*{0.5cm} c^-_A = \frac{ \alpha^- }{\sqrt{1 + |\alpha^-|^2}} 
\end{equation}
where the $c^{\pm}_B$ coefficients are found from using equation (\ref{Acoefficients}) and (\ref{ABcoefficients}).  

The magnetic susceptibility expressions require the evaluation of matrix elements of the $u_{n\textbf{k}}$. This is done by reducing these matrix elements to expressions in terms of the orthogonal atomic orbitals, and then evaluating those matrix elements explicitly by calculating the integrals. Matrix elements are defined as 
\begin{equation}
    \langle \hat{\mathcal{O}}(\textbf{x}) \rangle_{AB}^{\textbf{R}} = \int d\textbf{x} \Phi_{A}^*(\textbf{x}-\textbf{R}) \hat{\mathcal{O}}(\textbf{x}) \Phi_{B}(\textbf{x}), 
\end{equation}
and we obtain
expressions for equations (\ref{connection}), (\ref{curvature}), and (\ref{angularmomentum}), using equations (\ref{LCAO})  and (\ref{cellperiodicfunctions}):

\begin{equation}
\begin{split}
    -i\xi^\mu_{nm} = (c^n_{A})^* (\partial_\mu c^m_A) + (c^n_B)^*(\partial_\mu c^m_{B})
    \\
    -i(c^n_A)^* c^m_B \sum_{\textbf{R}=\textbf{R}_A-\textbf{R}_B} e^{-i\textbf{k}\cdot\textbf{R}} \langle x_\mu \rangle^\textbf{R}_{AB} 
    \\
    -i(c^n_{B})^* c^m_A \sum_{\textbf{R}=\textbf{R}_{B}-\textbf{R}_{A}} e^{-i\textbf{k}\cdot\textbf{R}}\langle x_{\mu} \rangle_{BA}^\textbf{R},
\end{split}
\end{equation}

\begin{widetext}

\begin{equation}
\begin{split}
    \sum_{s} \xi^\mu_{ns}\xi^\nu_{sm} = 
    (\partial_\mu c^n_{A})^* (\partial_\nu c^m_{A}) 
    + (\partial_\mu c^n_{B})^* (\partial_\nu c^m_{B}) 
    + (c^n_A)^*(c^m_A) \langle x_{\mu} x_{\nu} \rangle^{AA}_{0} 
    + (c^n_B)^* c^m_B \langle x_\mu x_\nu \rangle^{BB}_0
    \\
    -i(\partial_\mu c^n_A)^* c^m_B \sum_{\textbf{R}=\textbf{R}_A-\textbf{R}_B} e^{-i\textbf{k}\cdot\textbf{R}} \langle x_{\nu} \rangle_{AB}^{\textbf{R}}
    +i(c^n_B)^* \partial_\nu c^m_A \sum_{\textbf{R}=\textbf{R}_B - \textbf{R}_A} e^{-i\textbf{k}\cdot\textbf{R}} \langle x_\mu \rangle^\textbf{R}_{BA} 
    \\
    -i(\partial_\mu c^n_{B})^* c^m_{A} \sum_{\textbf{R}=\textbf{R}_B-\textbf{R}_A} e^{-i\textbf{k}\cdot\textbf{R}} \langle x_{\nu} \rangle^{\textbf{R}}_{BA}
    +i(c^n_A)^* \partial_\nu c^m_{B} \sum_{\textbf{R}=\textbf{R}_A-\textbf{R}_B} e^{-i\textbf{k}\cdot\textbf{R}} \langle x_{\mu} \rangle^{\textbf{R}}_{AB} 
    \\
    + (c^n_{A})^* c^m_{B} \sum_{\textbf{R}=\textbf{R}_A-\textbf{R}_B} e^{-i\textbf{k}\cdot\textbf{R}}\Big( \langle x_{\mu} x_{\nu}\rangle_{AB}^\textbf{R} - R_{\mu} \langle x_{\nu} \rangle_{AB}^{\textbf{R}} \Big)
    \\
    + (c^n_{B})^* c^m_A \sum_{\textbf{R}=\textbf{R}_B-\textbf{R}_A} e^{-i\textbf{k}\cdot\textbf{R}} \Big( \langle x_{\mu} x_{\nu}\rangle_{BA}^\textbf{R} - R_{\mu} \langle x_{\nu} \rangle_{BA}^{\textbf{R}} \Big),
\end{split}
\end{equation}

\begin{equation}
\begin{split}
    \int d\textbf{x} \frac{\partial u^\dag_{n\textbf{k}}}{\partial k_{\mu}} \frac{\partial H_\textbf{k}}{\partial k_\nu} u_{m\textbf{k}} = -t(\partial_\mu c^n_A)^*(\partial_\nu \gamma_\textbf{k} c^m_B) -t(\partial_\mu c^n_{B})^* \partial_\nu \gamma_\textbf{k} c^m_A 
    -it (c^n_B)^* \partial_\nu \gamma_\textbf{k} c^m_B \sum_{\textbf{R}=\textbf{R}_A-\textbf{R}_B} e^{-i\textbf{k}\cdot\textbf{R}} \langle x_{\mu} \rangle^{\textbf{R}}_{AB} 
    \\
    -it(c^n_{A})^* \partial_\nu c^m_{A} \sum_{\textbf{R}=\textbf{R}_{B}-\textbf{R}_A} e^{-i\textbf{k}\cdot\textbf{R}} \langle x_{\mu} \rangle_{BA}^{\textbf{R}}.
\end{split}
\end{equation}
\end{widetext}

Under the nearest neighbour assumption this means evaluating matrix elements when $\textbf{R}=0,\pm\ \bf{\delta_{i}}$. The nearest neighbour matrix elements of interest have the form
\begin{equation}
    \langle \hat{\mathcal{O}}(\textbf{x})\rangle_{AB}^{\textbf{R}_{AB}} = \int d\textbf{x} \Phi^{*}_{A}(\textbf{x}-\textbf{R}_{AB}) \hat{\mathcal{O}}(\textbf{x}) \Phi_B(\textbf{x}),
\end{equation}
and
\begin{equation}
    \langle \hat{\mathcal{O}}(\textbf{x})\rangle_{AB(0)}^{\textbf{R}_{AB}} = \int d\textbf{x} \phi^{*}_{A}(\textbf{x}-\textbf{R}_{AB}) \hat{\mathcal{O}}(\textbf{x}) \phi_{B}(\textbf{x}),
\end{equation}
where $\textbf{R}_{AB} = \textbf{R}_A-\textbf{R}_{B} = -\textbf{R}_{BA}$. By substituting in the definitions of the orbitals orthogonal to first order in $s$ ((\ref{AtomicOrbitalA}) and (\ref{AtomicOrbitalB})), the matrix elements are

\begin{equation}
\begin{split}
    \langle \hat{\mathcal{O}}(\textbf{x})\rangle^{0}_{AA} = \langle \hat{\mathcal{O}}(\textbf{x}) \rangle^{0}_{AA(0)} 
    -s \text{Re}\sum_{\textbf{R}_{AB}} \langle \hat{\mathcal{O}}(\textbf{x}-\textbf{R}_{AB})\rangle_{AB(0)}^{\textbf{R}_{AB}},
\end{split}
\end{equation}

\begin{equation}
\begin{split}
    \langle \hat{\mathcal{O}}(\textbf{x})\rangle_{BB}^0 = \langle \hat{\mathcal{O}}(\textbf{x})\rangle_{BB(0)}^0 - s\text{Re} \sum_{\textbf{R}_{AB}} \langle \hat{\mathcal{O}}(\textbf{x})\rangle_{AB(0)}^{\textbf{R}_{AB}}, 
\end{split}
\end{equation}

\begin{equation}
\begin{split}
    \langle \hat{\mathcal{O}}(\textbf{x})\rangle_{AB}^{\textbf{R}_{AB}} =& \langle \hat{\mathcal{O}}(\textbf{x})\rangle_{AB(0)}^{\textbf{R}_{AB}}
    \\
    &- \frac{s}{2} \langle \hat{\mathcal{O}}(\textbf{x})\rangle_{BB(0)}^{0} 
    -\frac{s}{2} \langle \hat{\mathcal{O}}(\textbf{x}+\textbf{R})\rangle_{AA(0)}^{0},
\end{split}
\end{equation}
where $\hat{\mathcal{O}}(\textbf{r})$ represents any of the relevant operators 
such as components of the position operator or products of them ($x$ or $xy$, etc.). It suffices to consider the coordinate directions perpendicular and parallel to the displacement \textbf{R} of the orbitals; these are calculated in closed form using a prolate spheroidal coordinate transformation \cite{AbramowitzStegun}. 
By a simple change of variables,
\begin{equation}
\begin{split}
    \langle x_{\mu}\rangle_{BA}^{\textbf{R}_{BA}} =& \langle x_{\mu} \rangle_{AB}^{\textbf{R}_{AB}} 
    \\
    \langle x_{\mu} x_{\nu} \rangle_{BA}^{\textbf{R}_{BA}} =& 
    \langle x_{\mu} x_{\nu} \rangle_{AB}^{\textbf{R}_{AB}} 
    + R_{BA \nu} \langle x_{\mu} \rangle_{AB}^{\textbf{R}_{AB}} 
    \\
    &+ R_{BA\mu} \langle x_{\nu} \rangle_{AB}^{\textbf{R}_{AB}}.
\end{split}
\end{equation}

\subsection{Complete Evaluation of $\chi_\text{inter}$}

This section is a direct generalization of the method applied by Ogata \cite{OgataMagSusIII}. By applying the order of $s$ assumption, $L_{z} \phi_{pz} = 0$, and writing $\frac{\partial H_{k}}{\partial k_\mu}$ directly as an operator in real-space coordinates acting on the wavefunctions we can write

\begin{equation}
\begin{split}
    &\frac{\partial H_\textbf{k} }{\partial k_y} \frac{ \partial u^{\pm}_\textbf{k}}{\partial k_x} - \frac{\partial H_\textbf{k}}{\partial k_x} \frac{\partial u^{\pm}_\textbf{k}}{\partial k_y} 
    \\
    &= \Big( \zeta^{\pm}_{1x} \frac{\partial H_\textbf{k}}{\partial k_x} + \zeta^{\pm}_{1y} \frac{ \partial H_\textbf{k}}{\partial k_y} \Big) u^{\pm}_\textbf{k} + \Big( \zeta^{\pm}_{2x} \frac{\partial H_\textbf{k}}{\partial k_x} + \zeta^{\pm}_{2y} \frac{\partial H_\textbf{k} }{\partial k_y} \Big) u^{\mp}_\textbf{k},
\end{split}
\end{equation}
where the $\zeta$'s are given below in equation (\ref{Zetas}). Then using the definition of the velocity matrix element,
\begin{equation}
    \frac{1}{\mathcal{V}_{uc}}\int_{\mathcal{V}_{uc}} d\textbf{x} u^\dag_{l\textbf{k}} \frac{\partial H_\textbf{k}}{\partial k_{\mu}} u_{l'\textbf{k}} = i(\epsilon_l - \epsilon_{l'}) \xi^\mu_{ll'} + \frac{\partial \epsilon_{l'}}{\partial k_\mu} \delta_{ll'},
\end{equation}
the non-Hermitian spontaneous magnetization matrix element can be evaluated as 
\begin{equation}
\begin{split}
    \mathcal{M}^{z}_{-,l} =
    -\frac{ie}{2\hbar c}\Bigg[ 
    2i\epsilon_{+} \zeta^{-*}_{2x} \xi^x_{+,l} 
    + 2i\epsilon_{+} \zeta^{-*}_{2y} \xi^y_{+,l}
    \\
    +i(\epsilon_{-} - \epsilon_l) \Big( \zeta^{-*}_{1x} \xi^x_{-,l} + \zeta^{-*}_{1y} \xi^{y}_{-,l}
    + \zeta_{2x}^{-*} \xi^x_{+,l} + \zeta^{-*}_{2y} \xi^y_{+,l} \Big)
    \\
    + \zeta^{-*}_{1x} \frac{\partial \epsilon_-}{\partial k_x} \delta_{l,-} 
    + \zeta^{-*}_{1y} \frac{\partial \epsilon_-}{\partial k_y} \delta_{l,-}
    + \zeta^{-*}_{2x} \frac{\partial \epsilon_+}{\partial k_x} \delta_{l,+}
    \\
    +\zeta^{-*}_{2y} \frac{\partial \epsilon_+}{\partial k_y} \delta_{l,+}
    +\frac{\partial \epsilon_-}{\partial k_x} \xi^y_{-,l} 
    - \frac{\partial \epsilon_-}{\partial k_y} \xi^x_{-,l}
    \Bigg].
\end{split}
\end{equation}
Since terms of the form $(\partial_\mu \epsilon )^2/(\epsilon_- - \epsilon_l)$ for $l\neq +$ are dropped as second order, the absolute square can be evaluated using the identities 

\begin{equation}
    \sum_{l' \neq -} \xi^\nu_{n,l'} \xi^\mu_{l',m} = \int d\textbf{x} \frac{\partial u^\dag_{n\textbf{k}}}{\partial k_{\nu}} \frac{\partial u_{m\textbf{k}}}{\partial k_{\mu}} - \xi^\mu_{n,-} \xi^\nu_{-,m}, 
\end{equation}
and
\begin{equation}
\begin{split}
    \sum_{l'\neq -} (\epsilon_{-} - \epsilon_{l'}) \xi^\nu_{n,l'} \xi^\mu_{l',m} 
    = \int d\textbf{x} \frac{\partial u^\dag_{n\textbf{k}}}{\partial k_\nu} (\epsilon_{-} - H_\textbf{k}) \frac{\partial u_{m\textbf{k}}}{\partial k_\mu} 
    \\
    = (\epsilon_{-} - \epsilon_m) \int d\textbf{x} \frac{\partial u^\dag_{n\textbf{k}}}{\partial k_\nu} \frac{\partial u_{m\textbf{k}}}{\partial k_{\mu}} 
    + \int d\textbf{x} \frac{\partial u^\dag_{n\textbf{k}}}{\partial k_{\nu}} \frac{\partial H_\textbf{k}}{\partial k_\mu} u_{m\textbf{k}}
    \\
    - i \frac{\partial \epsilon_m}{\partial k_{\mu}} \xi^\nu_{nm}.
\end{split}
\end{equation}

Note that in graphene the energy is directly proportional to $t$ by $\epsilon = t\gamma_\textbf{k}$, and since $t$ is an overlap integral, this is a first-order quantity in the sense of scheme II. Adding the bandgap $\Delta$ means that this proportionality no longer holds, so the energy in boron nitride is no longer strictly a first order quantity in the sense of scheme II. 
\begin{widetext}
The definitions of the four $\zeta$'s are 

\begin{equation}
\label{Zetas}
\begin{split}
    \zeta^{\pm}_{1x} = \frac{1}{c^\pm_A c^{\mp}_B 
    - c^{\mp}_A c^{\pm}_B} \Bigg( 
    c^{\mp}_{B}\Big( -\partial_y c^{\pm}_A + \frac{s}{2} c^{\pm}_B \partial_y \gamma_\textbf{k} \Big) 
    - c^{\mp}_A\Big(-\partial_y c^{\pm}_{B} +\frac{s}{2} c^{\pm}_A (\partial_y \gamma_\textbf{k})^* \Big)
    \Bigg)
    \\
    \zeta^{\pm}_{1y} = \frac{1}{c^\pm_A c^{\mp}_B 
    - c^{\mp}_A c^{\pm}_B} \Bigg( c_{B}^{\mp} \Big( \partial_x c^{\pm}_A - \frac{s}{2} c^{\pm}_B \partial_x \gamma_\textbf{k}\Big)
    - c^{\mp}_A \Big( \partial_x c^{\pm}_{B} - \frac{s}{2} c^{\pm}_A (\partial_x \gamma_\textbf{k})^*\Big)
    \Bigg)
    \\
    \zeta^{\pm}_{2x} = \frac{1}{c^\pm_A c^{\mp}_B 
    - c^{\mp}_A c^{\pm}_B} \Bigg( 
    -c^{\pm}_B \Big( -\partial_y c^{\pm}_A + \frac{s}{2} c^{\pm}_{B} \partial_y \gamma_\textbf{k} \Big) + c^{\pm}_A\Big( -\partial_y c^{\pm}_B + \frac{s}{2} c^{\pm}_A(\partial_y \gamma_\textbf{k})^* \Big)
    \Bigg) 
    \\
    \zeta^{\pm}_{2y} = \frac{1}{c^\pm_A c^{\mp}_B 
    - c^{\mp}_A c^{\pm}_B} \Bigg(
    -c_{B}^{\pm}\Big(\partial_x c^{\pm}_A - \frac{s}{2} c^{\pm}_B \partial_x \gamma_\textbf{k} \Big) + c_A^{\pm}\Big( \partial_x c^{\pm}_{B} - \frac{s}{2} c^{\pm}_A (\partial_x \gamma_\textbf{k})^* \Big)
    \Bigg)
\end{split}
\end{equation}

\end{widetext}
\bibliographystyle{apsrev4-2}
\bibliography{references.bib}

\begin{thebibliography}{47}
\bibitem [{Ear()}]{EarlyComparisons}%
  \BibitemOpen
  \href@noop {} {}\bibinfo {note} {In Roth's work they claim agreement with Hebborn and Sondheimer's formula for the case of no spin contributions, but offer no proof. Blount claims that the work of Adams, and Hebborn and Sondheimer are ``essentially equivalent formulas" without comment on the definition of essentially equivalent or proof.}\BibitemShut {Stop}%
\end{thebibliography}

\end{document}